\documentclass[journal]{IEEEtran}
\usepackage[utf8]{inputenc}
\usepackage{cite}

\usepackage{svg}
\usepackage{amsmath}
\usepackage{amssymb}
\usepackage{hyperref}
\usepackage{relsize}%
\usepackage{graphicx}
\usepackage{subfig}
\usepackage{xcolor}
\usepackage[utf8]{inputenc}
\usepackage[english]{babel}
\usepackage{amsthm}
\usepackage{xurl}
\newtheorem{remark}{Remark}
\usepackage{algorithmicx}
\usepackage[ruled,vlined, linesnumbered]{algorithm2e}
\usepackage{comment} 
\newcommand{\ignore}[1]{}
\usepackage{color} 
\usepackage{multirow}
\usepackage[T1]{fontenc}

\makeatletter
\newcommand{\removelatexerror}{\let\@latex@error\@gobble}

\IEEEoverridecommandlockouts
\begin{document}

\title{Deep Multi-agent Reinforcement Learning for Highway On-Ramp Merging in Mixed Traffic}

\author{Dong Chen$^{1}$, Mohammad R. Hajidavalloo$^{1}$, Zhaojian Li$^{*{1}}$,  Kaian Chen$^{1}$,\\ Yongqiang Wang$^{2}$, Longsheng Jiang$^{3}$, Yue Wang$^{3}$

\thanks{$^1$ Dong Chen, Mohammad Hajidavalloo, Zhaojian Li,  and Kaian Chen are with the Department of Mechanical Engineering, Michigan State University, Lansing, MI, 48824, USA. Email: {\tt\{chendon9, hajidava, lizhaoj1,  chenkaia\}@msu.edu.}}
\thanks{$^2$ Yongqiang Wang is with the Department of Electrical and Computer Engineering, Clemson University, Clemson, SC, 29630, USA. Email: {\tt{yongqiw@clemson.edu}}.}
\thanks{$^3$ Longsheng Jiang and Yue Wang are with the Department of Mechanical Engineering, Clemson University, Clemson, SC, 29634, USA. Email:  {\tt\{longshj,yue6\}@g.clemson.edu}.}

\thanks{$*$ Zhaojian Li is the corresponding author.}}

\maketitle


\begin{abstract}
On-ramp merging is a challenging task for autonomous vehicles (AVs), especially in mixed traffic where AVs coexist with human-driven vehicles (HDVs). In this paper, we formulate the mixed-traffic highway on-ramp merging  problem as a multi-agent reinforcement learning (MARL) problem, where the AVs (on both merge lane and through lane) collaboratively learn a policy to adapt to  HDVs to maximize the traffic throughput. We develop an efficient and scalable MARL framework that can be used in dynamic traffic where the communication topology could be time-varying. Parameter sharing and local rewards are exploited to foster inter-agent cooperation while achieving great scalability. An action masking scheme is employed to improve learning efficiency by filtering out invalid/unsafe actions at each step. In addition, a novel priority-based safety supervisor is developed to significantly reduce collision rate and greatly expedite the training process. A gym-like simulation environment is developed and open-sourced with three different levels of traffic densities. We exploit curriculum learning to efficiently learn harder tasks from trained models under simpler settings.   
Comprehensive experimental results show the proposed MARL framework consistently outperforms several state-of-the-art benchmarks.

\end{abstract}

\IEEEpeerreviewmaketitle

\section{Introduction}\label{sec:1}
Autonomous vehicle (AV) technologies, such as Tesla Autopilot \cite{autopilot} and Baidu Apollo\cite{apollo}, have already been deployed in (semi-)autonomous vehicles on real-world roads. Despite the great advances  over the past decade that have made this possible, the number of traffic accidents involving AVs are increasing in recent years \cite{dixit2016autonomous, favaro2017examining}. The accidents are often caused by the inability of AVs to timely react to the dynamic driving environment, especially in a mixed traffic with both AVs and human-driven vehicles (HDVs); the AVs need not only to react to road objects but also to attend to the behaviors of HDVs. 
Among the many challenging driving scenarios,  highway on-ramp merging is one of the most difficult tasks for AVs \cite{ni2005simplified, leclercq2011capacity}, which is the topic of this paper. 

\begin{figure}[]
  \centering
  \includegraphics[width=0.46\textwidth]{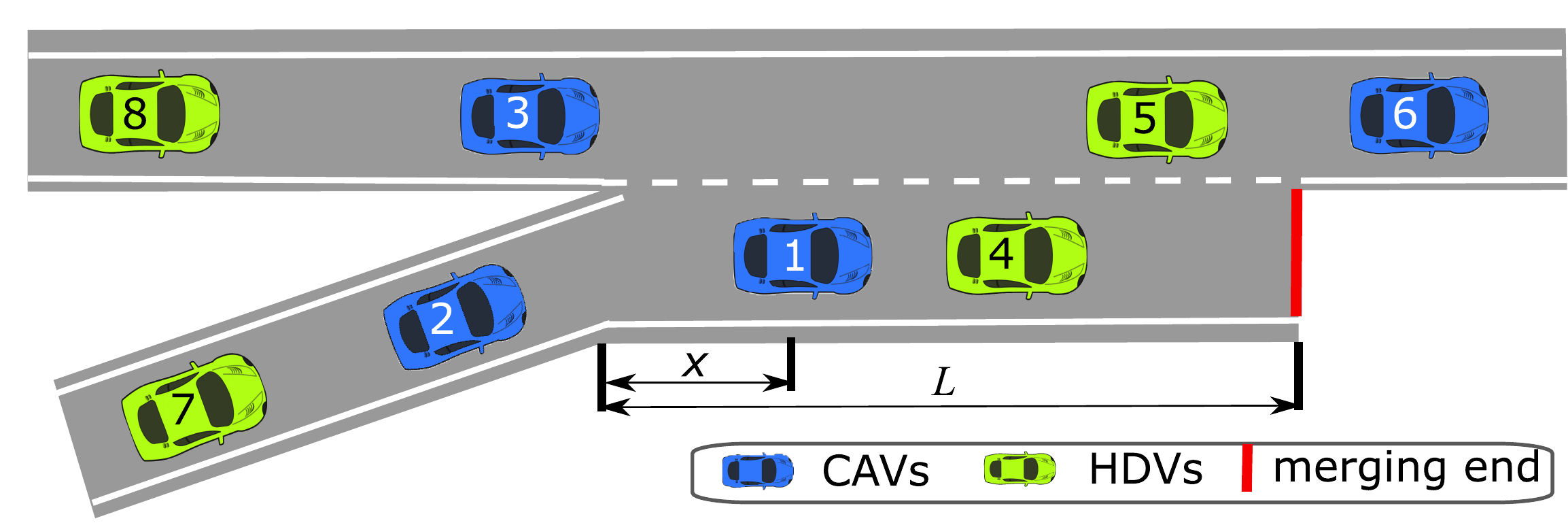}
  \caption{Illustration of the considered on-ramp merging traffic scenario. CAVs (blue) and HDVs (green) coexist on both ramp and through lanes. }
  \label{fig:merging_scenario}
\end{figure}

The considered on-ramp merging scenario is illustrated  in Fig.~\ref{fig:merging_scenario}, where we consider a general setup that AVs and HDVs coexist on both merge lane and through lane.  On-ramp vehicles need to efficiently merge onto the through lane without collision. In an ideal cooperative setting, the vehicles on the through lane should proactively  decelerate or accelerate to make adequate space for on-ramp vehicles to safely merge whereas   the on-ramp vehicles also adjust speed and promptly cut in when it is safe, to avoid  deadlock situations \cite{bouton2019cooperation}. It is clear that coordination between the vehicles is a crucial enabler for safe and efficient merging maneuvers. While this is relatively easy to achieve in a full-AV scenario, AV coordination in the presence of HDVs is a significantly more challenging task.

Rule-based and optimization-based methods have been proposed to tackle the automated merging problem \cite{rios2016survey,jacobson1989real, hourdakis2002evaluation, lin2019anti}. In particular, rule-based approaches employ heuristics and hard-coded rules to guide the AVs \cite{jacobson1989real, hourdakis2002evaluation}. While this is feasible for simple traffic scenarios, these methods quickly become impractical in more complex merging scenarios \cite{cao2015cooperative}. In an optimal control setting, vehicle interactions are modeled as a dynamic system with actions from controlled vehicles as inputs. For example, a model predictive control (MPC) approach is developed to control an AV to merge in a parallel-type ramp \cite{cao2015cooperative}. While promising results are demonstrated, the MPC-based methods rely on accurate dynamic merging models (including human driving models) and are typically computationally-involved as online optimizations are needed at each time step \cite{MPCbook}. Extensive surveys on model-based control strategies for on-ramp merging are presented in \cite{papageorgiou2003review, papageorgiou2002freeway, papamichail2008traffic}. However, those approaches only consider fully automated vehicles, which are not applicable to the considered mixed-traffic setting.

On the other hand, data-driven methods such as reinforcement learning (RL) have received increased attention and been
explored for AV highway merging \cite{lin2019anti, lubars2020combining}. Specifically, a multi-objective reward function for safety and jerk minimization is designed for AV merging and the Deep Deterministic Policy Gradient (DDPG) algorithm \cite{lillicrap2015continuous} is exploited to solve the RL problem in \cite{lin2019anti}. In \cite{lubars2020combining},  RL and MPC are integrated to promote the learning efficiency, which achieves a good trade-off between passenger comfort, efficiency, crash rate, and robustness. However, those approaches are only designed for a single AV, treating all other vehicles as part of the environment.

In this paper, we treat a general setup (see Fig.~\ref{fig:merging_scenario}) where multiple AVs learn to adapt to HDVs and cooperatively accomplish merging tasks to maximize traffic throughput safely. As a result, it is natural to extend the single-agent RL to a multi-agent reinforcement learning (MARL) framework where the AVs collaboratively learn control policies to achieve the aforementioned goal (see Section II.B for a review of state-of-the-art MARL algorithms). However, this is a challenging task due to dynamic connectivity topology, complex motion patterns involving AV coupled dynamics, and intricate decision makings. This complexity is even more pronounced when human drivers are involved.  

While several MARL approaches have been developed for connected and autonomous vehicles (CAVs)  in car-following and lane overtaking scenarios \cite{ wangmulti, palanisamy2020multi, kaushik2018parameter, ha2020leveraging, bhalla2020deep, dong2020drl, yu2019distributed},  to the best of our knowledge, no MARL algorithm has been developed  for  the considered highway on-ramp merging scenario. 
In this work, we develop a novel decentralized MARL framework to enable AVs to efficiently learn a safe and efficient policy in the highway on-ramp merging scenario where a general policy is learned for vehicles on both lanes.  A priority-based safety supervisor is designed to enhance safety and improve learning efficiency, through sequential and multi-step predictions. Parameter sharing and local rewards are exploited to foster inter-agent cooperation while achieving great  scalability. The main contributions and the technical  advancements of this paper are summarized as follows.

\begin{enumerate}

\item We formulate the mixed-traffic on-ramp merging problem (with AVs and HDVs coexisting on both ramp and through lanes) as a decentralized MARL problem. The formulation can allow for a dynamic environment with a time-varying connectivity topology. A corresponding gym-like simulation platform with three different levels of traffic density is developed and open-sourced\footnote{See \url{https://github.com/DongChen06/MARL_CAVs}}.

\item  We develop a novel, efficient, and scalable MARL algorithm, featuring a parameter-sharing mechanism, effective reward function design, and action masking.  Furthermore, a priority-based safety supervisor is developed, which significantly reduces collision rates in training and subsequently improves learning efficiency.

\item We employ curriculum learning to speed up the learning for harder tasks by building upon trained models from less complex traffic scenarios. 

\item We conduct comprehensive experiments, and the results show that the proposed approach consistently outperforms several state-of-the-art algorithms in terms of driving safety and efficiency. 
\end{enumerate}

The remainder of the paper is organized as follows. Section~\ref{sec:2} briefly introduces  RL and MARL, and reviews state-of-the-art algorithms. The problem formulation and the proposed MARL framework are described in Section~\ref{sec:3} whereas the priority-based safety supervisor is detailed in Section~\ref{sec:4}. Experiments, results, and discussions are presented in Section~\ref{sec:5}. We conclude the paper and discuss  future works in Section~\ref{sec:6}.

\section{Background}\label{sec:2}
In this section, we review the preliminaries of RL and introduce several state-of-the-art MARL algorithms to put our proposed work in proper context. Readers who are familiar with the RL and MARL literature can skip this section and jump to Section III directly.

\subsection{Preliminaries of Reinforcement Learning (RL)}
In a RL setting, at each time step $t$, the agent observes the state $s_t \in \mathcal{S} \subseteq \mathbb{R}^{n}$, takes an action $a_t \in \mathcal{A} \subseteq \mathbb{R}^{m}$,  and subsequently receives a reward signal $r_t \in \mathbb{R}$ and an updated state $s_{t+1}$ at time $t+1$ from the environment.  The goal of the RL agent is to learn an optimal policy $\pi^{*}: \mathcal{S} \rightarrow \mathcal{A}$, a mapping from state to action, that maximizes the accumulated reward $R_t = \sum_{k=0}^{T} \gamma ^k r_{t+k}$, where $r_{t+k}$ is the reward at time step $t+k$ and $\gamma\in (0,1]$ is the discount factor that quantifies the relative importance one wants to place on future rewards.

The state-action value function (or Q-function) under policy $\pi$, denoted by $Q^{\pi}(s_t,a_t)$, is an estimation of the expected return (accumulated reward in an infinite horizon) if starting from state $s_t$, taking an immediate action $a_t$, and then following policy $\pi$ afterwards. The optimal Q-function can be characterized by the following Bellman equation, $Q^{*}(s_t,a_t) = E [r(s_t, a_t) + \gamma  \sum_{s_{t+1}} P(s_{t+1} |s_t, a_t) \max_{a_{t+1}} Q^{*}(s_{t+1},a_{t+1})]$, where the next state $s_{t+1}$ is sampled from the environment’s transition rules $P(s_{t+1} |s_t, a_t)$.
The state value function of a state $s_t$ under policy $\pi$, $V^{\pi}(s_t)$, is defined as the expected return if starting from $s_t$ and immediately following policy $\pi$, i.e., $V^{\pi}(s_t) = E_{\pi}{[R_t | {s_t = s}]}$.  Often the agent's policy is parameterized by some parameters $\theta$ and the goal is to learn appropriate $\theta$ to achieve desired system behavior. In actor-critic (A2C) algorithms \cite{mnih2016asynchronous}, two networks are employed: a critic network parameterized by $\phi$ to learn the value function $V_{\phi}^{\pi_\theta}(s_t)$ and an actor network $\pi_{\theta}(a_t|s_t)$ parameterized by $\theta$. The policy network is updated by maximizing the following objective function:
\begin{equation}\label{eqn:advantagepolicygradient}
J^{\pi_{\theta}} = E_{\pi_{\theta}} \left[ \log \pi_{\theta}(a_t|s_t)
 A_t \right],
\end{equation}
where $A_t= Q^{\pi_\theta}(s_t,a_t) - V_{\phi}(s_t)$  is the advantage function  that characterizes the improvement on reward if taking action $a_t$  over the average reward of  all possible actions taken at state $s_t$ \cite{mnih2016asynchronous}. The value function parameter $\phi$ is updated by minimizing the following loss function: 
\begin{equation}\label{eqn:valueloss}
J^{V_\phi} = \min_{\phi} E_{\mathcal{D}}\Big (R_t + \gamma V_{\phi'} (s_{t+1}) - V_{\phi}(s_t)\Big )^2,
\end{equation}
where $\mathcal{D}$ denotes an experience replay buffer that collects previously encountered experiences and $\phi'$ denotes the parameters obtained from earlier iterations used in a target network \cite{mnih2013playing}.

\subsection{Multi-agent Reinforcement Learning (MARL)}
MARL has found great successes across a wide
range of multi-agent systems, including traffic light control\cite{chu2019multi}, games \cite{berner2019dota}, resource management in wireless networks \cite{naderializadeh2021resource}, and powergrid control \cite{chen2020powernet}, only to name a few. MARL algorithms can be categorized into two main classes: {\it cooperative} and {\it non-cooperative}. In this paper, we will focus on the {\it cooperative} setting where all agents are encouraged to cooperate to achieve a common goal, i.e., safely maneuver with maximum traffic efficiency. We next introduce a few state-of-the-art cooperative MARL algorithms that we will use as benchmarks for comparison in Section~V. 

An independent MARL framework, called IQL,  is proposed in \cite{tan1993multi}, allowing each agent to learn independently and simultaneously while viewing other agents as part of the environment. While fully scalable, it suffers from non-stationarity and partial observability. An off-policy MARL algorithm is proposed in \cite{lowe2017multi} where collaboration is achieved by estimating the state-action value function using a centralized critic network based on global observations and actions. 
In \cite{chu2019multi}  a learnable communication protocol and a spatial distance factor are proposed to scale down the reward signals of neighboring agents during training. Experimental results show good scalability and improved cooperation among agents. However, these MARL approaches only consider a stationary environment with fixed communication topology and thus the algorithms need to be re-designed and/or re-trained whenever the communication typology changes.

Recently, parameter sharing  is widely applied in MARL settings with homogeneous agents \cite{kaushik2018parameter, lin2018efficient, terry2021revisiting}, which bootstraps single-agent RL methods and learns an identical policy for each agent, and thus enables the handling of changes in the number of participating agents. In \cite{terry2021revisiting}, several state-of-the-art single RL algorithms (i.e.,  PPO \cite{schulman2017proximal} and ACKTR\cite{wu2017scalable}) are extended to the MARL with parameter sharing denoted as MAPPO and MAACKTR. A parameter sharing A2C (MA2C) algorithm is proposed in \cite{lin2018efficient} to solve the fleet management problem and experimental results are given to confirm the performance. These methods will be used as benchmarks for performance comparison in Section~V. 

Several recent works also address safety issues in MARL problems. For example,   a centralized shielding approach is introduced in \cite{elsayed2021safe}, where a centralized model is used to monitor the joint actions of all agents and restrict unsafe actions. To address the scalability problem of centralized supervision, a local shielding approach is developed for only a subset of agents. Experiments in two-player navigation games in the grid world show good performance on collision avoidance. 
In addition, \cite{cai2021safe} proposes a decentralized control barrier function which shields unsafe actions based on available local information.  They demonstrate the performance of proposed approach using patrol tasks where two agents navigate in an environment with obstacles and walls. However, these methods consider an all-autonomous, cooperative agent environment,  without considering moving objects like HDVs.

To fill the aforementioned gaps, in this paper, we develop a novel on-policy MARL algorithm for the considered on-ramp merging problem with great efficiency and safety, which features action masking, priority-based safety supervisor, parameter sharing, and local reward shaping. Performance comparison between the proposed algorithm and the above benchmarks are presented in Section~V.

\section{Ramp Merging as MARL}\label{sec:3}
In this section, we first formulate the considered on-ramp merging problem as a partially observable Markov decision process (POMDP) \cite{hausknecht2015deep}. Then we present our actor-critic-based MARL algorithm, featuring a parameter-sharing mechanism, effective reward function design, and action masking, to solve the formulated POMDP, which is denoted as the baseline method in Section~V.

\subsection{Problem Formulation}\label{problem_formalation}
In this paper, we model the on-ramp merging environment in a mixed traffic as a model-free multi-agent network  \cite{chu2019multi,chen2020powernet}, $\mathcal{G} = (\text{\larger[2]$\nu$}, \text{\larger[2]$\varepsilon$})$, where each agent $i \in \text{\larger[2]$\nu$}$  communicates with its neighbors $\mathcal{N}_i:=\{j|\varepsilon_{ij}\in\large\text{\larger[2]$\varepsilon$})\}$ through the edge connections $\text{$\varepsilon_{ij}$}, i \neq j$. 
Let $S := \times_{i \in \text{\larger[2]$\nu$}} S_i$ and $\mathcal{A} := \times_{i \in \text{\larger[2]$\nu$}} \mathcal{A}_i$ denote the global state space and action space, respectively. The underlying dynamics of the system can be characterized by the state transition distribution $\mathcal{P}$: $\mathcal{S} \times \mathcal{A} \times \mathcal{S} \rightarrow [0, 1]$. We consider a decentralized MARL framework  where each agent $i$ (AV $i$) only observes a part of the environment (i.e., surrounding vehicles). This is consistent with the reality that AVs can only sense or communicate with vehicles in the close vicinity, making the overall dynamical system a POMDP $\mathcal{M_G}$, which can be described by the following tuple $(\{\mathcal{A}_i, \mathcal{S}_i, \mathcal{R}_i\}_{i\subseteq \nu}, \mathcal{T})$: 
\begin{figure}[!h]
  \centering
  \includegraphics[width=0.45\textwidth]{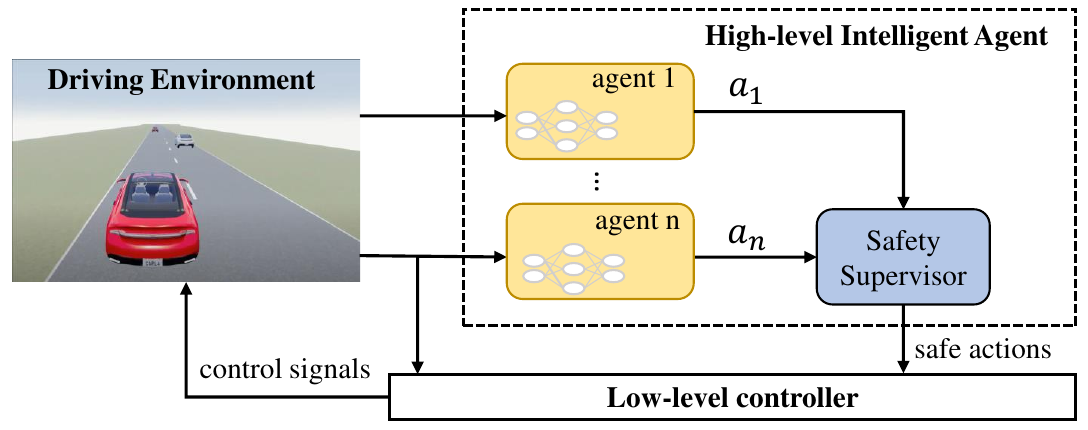}
  \caption{Schematics of system and simulation setup.}
  \label{fig:framework}
\vspace{-10pt}
\end{figure}

\begin{itemize}
\item \textbf{Action Space}: The action space $\mathcal{A}_i$ of agent $i$ is defined as the set of high-level control decisions, including {\it turn left, turn right, cruising, speed up,} and  {\it slow down} following the designs in \cite{li2017explicit, chen2020autonomous}. With a selected high-level decision,  lower-level controllers will then produce the corresponding steering and throttle control signals to maneuver the AVs. The system and simulation setup is illustrated in Fig.~\ref{fig:framework}. The overall action space of the system is the joint actions from all AVs, i.e., $\mathcal{A} =\mathcal{A}_{1}\times \mathcal{A}_{2}\times \cdots \times \mathcal{A}_{N}$.

\item \textbf{State Space}: The state of agent $i$, $\mathcal{S}_i$, is defined as a matrix of dimension $N_{\mathcal{N}_i} \times W$, where $N_{\mathcal{N}_i}$ is the number of observed vehicles and $W$ is the number of features used to represent the state of a vehicle, including:
\begin{itemize}
    \item \textit{ispresent}: a binary variable denoting whether a vehicle is observable in the vicinity of the ego vehicle.
    \item \textit{$x_l$}: the longitudinal position of the observed vehicle relative to the ego vehicle.
    \item \textit{$y$}: the lateral position of the observed vehicle relative to the ego vehicle.
    \item \textit{$v_x$}: the longitudinal speed of the observed vehicle relative to the ego vehicle.
    \item \textit{$v_y$}: the lateral speed of the observed vehicle relative to the ego vehicle.
\end{itemize}
We assume that only the ``neighboring vehicles'' can be observed by the ego vehicle.
The ``neighboring vehicles'' are defined as the nearest $N_{\mathcal{N}_i}$ vehicles that are within a $150~m$ longitudinal distance from the ego vehicle due to the local observability \cite{yu2019distributed}. In the considered on-ramp merging case as shown in Fig.~\ref{fig:merging_scenario}, we found $N_{\mathcal{N}_i} = 5$ achieves the best performance. The entire state of the system is then the Cartesian product of the individual states, i.e.,  $\mathcal{S} =\mathcal{S}_1\times \mathcal{S}_{2}\times \cdots\times \mathcal{S}_N$.

\item \textbf{Reward Function}: The reward function $\mathcal{R}_i$ is crucial to train the RL agents  so that it follows desired behaviors. As the objective is to train our agents to safely and efficiently pass the merging area, the reward for the $i$th agent at time step $t$ is defined as follows:
\begin{equation}\label{eqn:reward_function}
r_{i,t} = w_c r_c + w_s r_s + w_h r_h + w_m r_m,
\end{equation}
where $w_c$, $w_s$, $w_h$, and $w_m$ are positive weighting scalars corresponding to collision evaluation $r_c$, stable-speed evaluation $r_s$, headway time evaluation $r_h$, and merging cost evaluation $r_m$, respectively.
As safety is the most important criteria,  we make $w_c$ much bigger than other weights to prioritize safety.
The four performance metrics are defined as follows:
\begin{itemize}
\item the collision evaluation $r_c$ is set to -1 if collision happens, otherwise $r_c = 0$.
\item the speed evaluation $r_s$ is defined as 
    \begin{equation}\label{eqn:reward_speed}
        r_s = \min\left\{\frac{v_t - v_{min}}{v_{max} - v_{min}},\,1\right\},
    \end{equation}
    where $v_t$ is the current speed of the ego vehicle. Combining the speed recommendation from the US Department of Transportation (20-30 $m/s$ \cite{transportation2011policy}) and the speed range observed in the Next Generation Simulation (NGSIM) dataset\footnote{https://ops.fhwa.dot.gov/trafficanalysistools/ngsim.htm} (minimum speed at 6-8 $m/s$ \cite{thiemann2008estimating}), we set the minimum and maximum speeds of the ego vehicle as $v_{min}= 10~m/s$, and $v_{max}= 30~m/s$, respectively.
    
\item the time headway evaluation is defined as: 
    \begin{equation}\label{eqn:reward_headway}
        r_h = \log {\frac{d_{\text{headway}}}{t_h v_t}},
    \end{equation}
    where $d_{headway}$ is the distance headway  and $t_h$ is a predefined time headway threshold. As such, the ego vehicle will get penalized when the time headway is less than $t_h$ and rewarded only when the time headway is greater than $t_h$. In this paper, we choose $t_h$ as $1.2~s$ as suggested in \cite{ayres2001preferred}.

\item The merging cost $r_m$ is designed to penalize the waiting time on the merge lane to avoid deadlocks \cite{bouton2019cooperation}. Here we adopt $r_m = -\exp ({-(x-L)^2} / {10 L})$, where $x$ is the distance the ego vehicle has navigated on the ramp  and $L$ is the length of the ramp (see Fig.~1). The  merging cost function is plotted in Fig.~\ref{fig:merging_cost}, which shows that the penalty increases as the ego vehicle moves closer to the merging end.

\begin{figure}[!ht]
  \centering
  \includegraphics[width=0.44\textwidth]{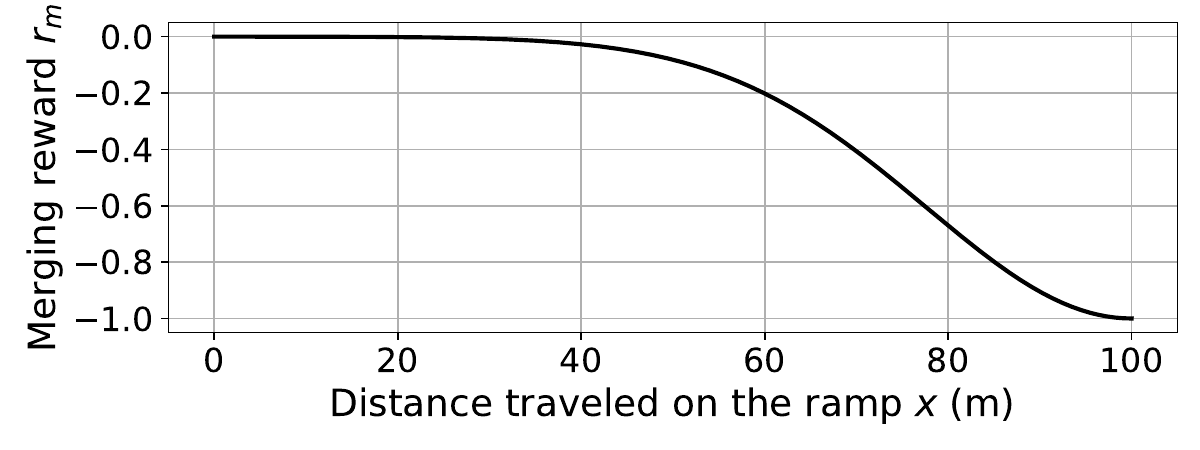}
  \caption{Illustration of the designed merging reward/penalty.}
  \label{fig:merging_cost}
\end{figure}
\end{itemize}

\item \textbf{Transition Probabilities}: the transition probability $\mathcal{T}(s'|s,a)$ characterizes the dynamics of the system. In the developed simulator, we exploit the intelligent driver model (IDM) \cite{treiber2000congested} and MOBIL model \cite{kesting2007general} for longitudinal acceleration and lateral lane change decisions of HDVs, respectively. The high-level decisions of AVs are made by the MARL algorithm and will be tracked by the lower-level controller (PID controller) (see Fig.~\ref{fig:framework}). A kinematic bicycle model \cite{polack2017kinematic} is used to propagate vehicle trajectories. 
We do not assume any prior knowledge of the transition probability in the development of our MARL algorithm. 
\end{itemize}

\subsection{MA2C for CAVs}
In the cooperative MARL setting, the objective is to maximize the global reward $R_{g,t} = \sum_{i=1}^{N} r_{i,t}$. Ideally, each agent will be assigned with the same average global reward $R_{t} = \frac{1}{N} R_{g,t}$ during training, i.e., $r_{1, t} = r_{2, t} = \cdots =r_{N, t}$. However,  this shared reward approach does not accurately indicate the contributions of each vehicle and can lead to several issues \cite{wolpert2002optimal, wang2020shapley}. First, aggregating the global reward can cause large latency and increase the communication overheads, which is problematic for systems with real-time constraints such as AVs. Second, a single global reward leads to the credit assignment problem~\cite{sutton2018reinforcement}, which can significantly impede the learning efficiency and limit the number of agents to a small size.
Therefore, in this paper, we adopt a local reward assignment strategy, where each ego vehicle is only affected by its neighboring vehicles. Specifically, the reward for the $i$th agent at time $t$ is defined as:
\begin{equation}\label{eqn:local_reward}
        r_{i, t} = \frac{1}{|{\nu_i}|} \sum_{j\in\nu_i} r_{j,t},
\end{equation}
where $\nu_i= {i}\cup\mathcal{N}_i$ is a set containing the ego vehicle and its neighbors, and $|\cdot|$ denotes the cardinality of a set. This local reward design only includes  rewards from agents that are most related to the success or failure of a task \cite{lin2018efficient, bagnell2005local}. This is appropriate for on-road vehicles as a vehicle only interacts with its surrounding vehicles and distant vehicles have limited impact on the ego vehicle.

\begin{figure*}[!ht]
  \centering
  \includegraphics[width=0.74\textwidth]{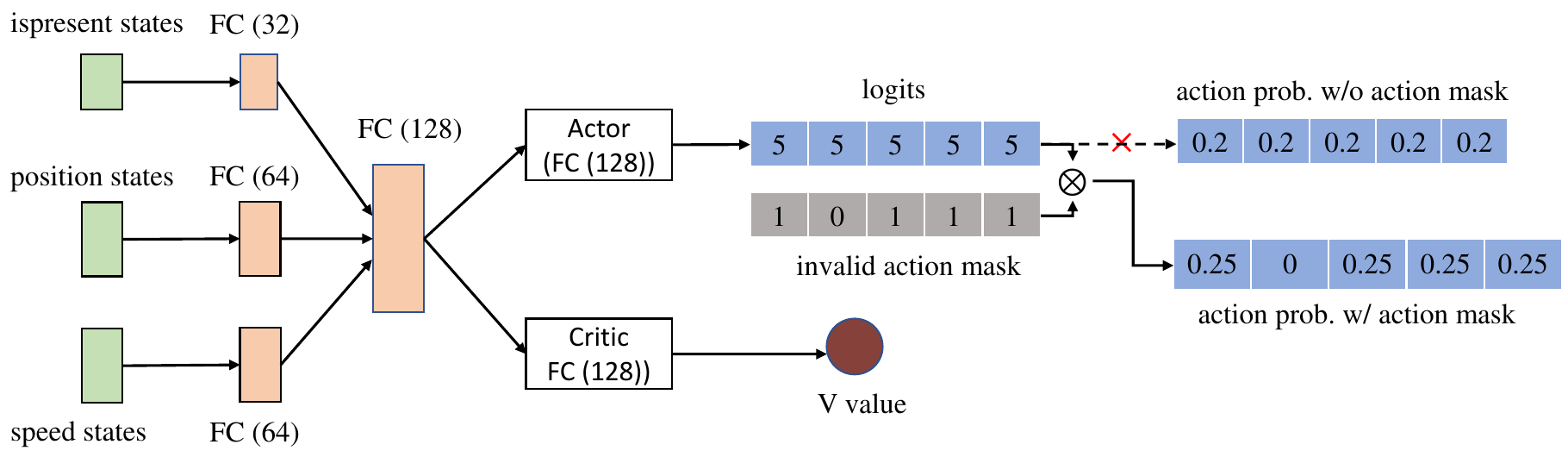}
  \caption{Architecture of the proposed network structure. The numbers in the parentheses denote the size of the layers. ``w/o" and ``w/" represent ``without" and ``with", respectively.}
  \label{fig:network_sturcture}
\end{figure*}

The used network backbone is shown in Fig.~\ref{fig:network_sturcture}, where the actor network and the critic network share the same low-level representations, and the policy loss and the value function error loss are thus combined into a single loss function \cite{schulman2017proximal}. With the shared network parameters, the overall loss function takes the following form:
\begin{equation}\label{eqn:objective_fn}
J(\theta_i) = J^{\pi_{\theta_i}} -  \beta_1 J^{V_{\phi_i}} + \beta_2 H(\pi_{\theta_i}(s_t)),
\end{equation}
where $\beta_1$ and $\beta_2$ are the weighting coefficients for the value function loss and the entropy regularization term, $H(\pi_{\theta_i}(s_t)) = E_{\pi_{\theta_i}} \left[- \log(\pi_{\theta_i}(s_t))\right]$,  used to encourage the agents to explore new states \cite{williams1992simple, schulman2017proximal}, respectively.
From Eq.~\ref{eqn:advantagepolicygradient}, it follows that the policy loss can be written as:
\begin{equation}\label{eqn:advantagepolicygradient_MARL}
J^{\pi_{\theta_i}} = E_{\pi_{\theta_i}} \left[ \log \pi_{\theta_i}(a_{i, t}|s_{i, t})
 A^{\pi_{\theta_i}}_{i, t} \right],
\end{equation}
where $A^{\pi_{\theta_i}}_{i, t} = r_{i,t} + \gamma V^{\pi_{\phi_i}}(s_{i, t+1}) - V^{\pi_{\phi_i}}(s_{i, t})$ is the advantage function  and $V^{\pi_{\phi_i}}(s_{i,t})$ is the state value function.
The loss for updating the state value $V_{\phi_i}$ is in the following form:
\begin{equation}\label{eqn:valueloss_MARL}
J^{V_{\phi_i}} = \min_{\phi_i} E_{\mathcal{D}_i} \Big [r_{i, t} + \gamma V_{\phi_i}(s_{i, t+1}) - V_{\phi_i}(s_{i, t})\Big ]^2.
\vspace{-5pt}
\end{equation}
We use separate experience reply buffers for each agent but the same policy network is updated with the same network parameters among agents. This is suitable as we train a general policy for both on-ramp and through AVs \cite{kaushik2018parameter, lin2018efficient}.
Minibatches of sampled trajectories are exploited to update the network parameters using Eq.~\ref{eqn:objective_fn} to reduce the variance.

\subsection{DNN Settings}
The deep neural network design is illustrated in Fig.~\ref{fig:network_sturcture}. Specifically, to improve scalability and robustness, we regroup the observation $s_{i, t}$ according to their physical units. For instance, the observation $s_{i,t}$ is divided into three groups: ${s}_{i,t}^1  \cup {s}_{i,t}^2 \cup {s}_{i,t}^3 $ according to their units, representing \textit{ispresent states}, \textit{position states} and \textit{speed states}, respectively. Each of the three sub-state vectors is encoded by one fully connected (FC) layer and the three encoded states are then concatenated into a single vector. The concatenated vector  is fed into a 128-neuron FC layer, the result of which is consumed by both the actor network and the critic network. In a standard setting, the logits $l_i$ from the actor network will go to a Softmax layer, producing the probability by ${\pi}_{\theta_i}(s_i) = \text{softmax} ([l_1, l_2, l_3, l_4, l_5])$, which is used to sample the actions, i.e., $a_i \sim \pi_{\theta_i}(s_i)$. 

However, this sampling procedure has several issues. First, invalid/unsafe actions are also assigned with non-zero probabilities; as a stochastic policy is used, these unsafe actions may be sampled during training, which can lead to undesirable system behaviors and even system breakdown. Second, sampling invalid/unsafe actions also impedes policy training as invalid policy updates \cite{huang2020closer} are executed for invalid actions, since the collected experiences associated with the invalid actions are not meaningful and misleading. To address these issues, we adopt the invalid action masking approach \cite{lin2018efficient} which ``masks out'' invalid actions and only samples from valid actions. As shown in Fig.~\ref{fig:network_sturcture}, with an invalid action mask obtained from the environment (e.g., based on the traffic scenario) where ``0'' represents an invalid action and ``1'' denotes a valid action, the corresponding logits of invalid actions are replaced with large negative values, e.g., $-1e8$. As a result, the probability of the invalid actions after the Softmax layer is very close to 0, and sampling from invalid actions can thus be avoided, equivalently ``renormalizing the probability distribution'' \cite{terry2021revisiting}. In this work, we consider the following invalid actions:
\begin{itemize}
\item the ego vehicle attempts to make lane changes to a non-existing lane. For example, the ego vehicle tries to make a left turn when it is already on the leftmost lane.
\item the ego vehicle attempts to speed up or slow down when its speed has already reached the predefined maximum or minimum speed.
\end{itemize}
Note that we only include the two most basic invalid actions here and other unsafe actions will be further checked and regulated by the proposed priority-based safety supervisor in Section~IV.

\section{Priority-based Safety Enhancement}\label{sec:4}
While obvious invalid actions can be avoided using the rule-based action masking scheme described above, it cannot prevent inter-vehicle or vehicle-obstacle collisions. Therefore, a more comprehensive safety supervisor is needed to deal with collisions in complex, dynamic, and cluttered mix-traffic environments. Towards that end, we propose a new safety-enhancement scheme by exploiting vehicle dynamics and multi-step predictions. The goal is to predict any potential collisions over a prediction horizon $T_n$ and correct the unsafe (exploratory) actions accordingly. As we consider a mixed traffic with HDVs, a proper model is needed to predict the high-level decisions of human drivers. In this paper, we use IDM \cite{treiber2000congested} to predict the longitudinal acceleration of the HDVs, based on the current speed and distance headway. In addition, we exploit the MOBIL lane change model \cite{kesting2007general} to predict the lane-changing behavior of HDVs,  which makes a lane-changing decision when it is safe and there is an extra acceleration gain. The high-level decisions of AVs  are generated by the MARL agent with the actions defined in Section~\ref{problem_formalation}.  These high-level acceleration and lane-change decisions will be realized through low-level PID controllers. The vehicle trajectories are then propagated based on the kinematic bicycle model \cite{polack2017kinematic}. We call the high-level decision-induced trajectories motion primitives and show the described framework and simulation setup in Fig.~\ref{fig:framework}. 

\subsection{Priority Assignment}
With the HDV motion models, one can predict whether a collision can happen in the next $T_n$ steps based on the joint motion primitives from all AVs. Therefore, it is attempting to use the joint action from all AVs to design the safety-enhancement scheme. However, while it is relatively straightforward to determine whether collisions can happen given a joint action, it is very computationally costly to determine a joint safe action, if a collision is detected, as the action space is $|\mathcal{A}_i|^N$ with $N$ being the number of AVs. It quickly becomes computationally intractable as $N$ grows, especially considering that the considered application has stringent real-time constraints. As such, we propose a sequential, priority-based safety enhancement scheme that has great computation efficiency and is thus suitable for real-time implementations. The principle is that we coordinate the AVs in a sequential order, prioritizing AVs with smaller safety margins.  For example, AVs near the end of the merge lane or near the defined safety boundary (e.g. distance headway very close to the defined threshold) should have higher priorities.  

More specifically, the following rationales are considered for priority assignments:
\begin{enumerate}
\item Vehicles on the merge lane should have higher priorities compared to vehicles on the through lane as vehicles on the merge lane face a time-critical merging task (due to merging lane end).
\item Merging vehicles closer to the merge lane end should have higher priorities as they are more probable to cause collisions and deadlocks \cite{bouton2019cooperation}.
\item Vehicles with smaller time headway should have a higher priority as they are more likely to collide with the preceding vehicles.
\end{enumerate}

Based on the above rationales, we construct the priority index $p_i$ of the ego vehicle $i$ as follows:
\begin{equation}\label{eqn:priority_coefficient}
p_i = \alpha_1 p_m + \alpha_2 p_d + \alpha_3 p_h + \sigma_i,
\vspace{0pt}
\end{equation}
where $\alpha_1$, $\alpha_2$ and $\alpha_3$ are positive weighting factors for the merging priority metric $p_m$, distance-to-end metric $p_d$, and time headway metric $p_h$, respectively; and $\sigma_i \sim \mathcal{N}(0,0.001)$ is a small random variable introduced to avoid two vehicles having the same priority indices.  Specifically, $p_m$ is defined as:
\begin{equation}
    p_m = 
    \begin{cases}
          0.5, & \text{if on merge lane}; \\
          0, &  \text{otherwise},
    \end{cases}
\end{equation}
which assigns priority score to vehicles on the merge lane. The distance-to-end priority score $p_d$ is defined as:
\begin{equation}
    p_d = 
    \begin{cases}
          \frac{x}{L}, & \text{if on merge lane}; \\
          0, & \text{otherwise},
    \end{cases}
\end{equation}
where $x$ and $L$ are the distance the ego vehicle has traveled on the connecting ramp and the length of the ramp (see Fig.~\ref{fig:merging_scenario}), respectively. Finally, we define the time headway priority score  to measure the headway priority as $p_h = - \log \frac{d_{\text{headway}}}{t_h v_t}$, where we use the time headway definition in Eq.~\ref{eqn:reward_headway}.

\subsection{Priority-based Safety Supervisor}
In this subsection, we present the proposed priority-based safety supervisor. Specifically, at each time step $t$, with the predicted HDV motions and assigned priority scores for all AVs as discussed above, the safety supervisor first generates a priority list for the AVs, $\mathcal{P}_t$, with their priority scores in a descendant order, i.e.,  vehicle with the highest priority is on top of the list.  Then the AV on the top of the obtained list, indexed by $\mathcal{P}_t[0]$, is selected for safety check. More specifically, based on  the (exploratory) action generated from the action network of vehicle $\mathcal{P}_t[0]$, the safety supervisor will examine whether the motion primitive induced by the exploratory action will conflict with its neighboring vehicles $\mathcal{N}_{P_t[0]}$ (both AVs and HDVs) in a considered time horizon $T_n$, where $T_n$ is a hyper-parameter that can be tuned. The motions of HDVs are predicted using the human-driver decision models and vehicle kinematic model discussed above, whereas the motions of all other (lower-priority) AVs are predicted assuming same actions from the last step. As the predicted trajectories of the considered vehicles (i.e., $\mathcal{P}_t[0]\cup\mathcal{N}_{\mathcal{P}_t[0]}$) are all $T_n$-step sequences, a collision can be detected if any two sequences have a distance below a prescribed threshold at any step $k$, $k=1,\cdots,T_n$. If no collision is predicted, the exploratory action will be chosen as the actual action for vehicle $P_t[0]$.

\begin{figure}[!ht]
  \centering
  \includegraphics[width=0.49\textwidth]{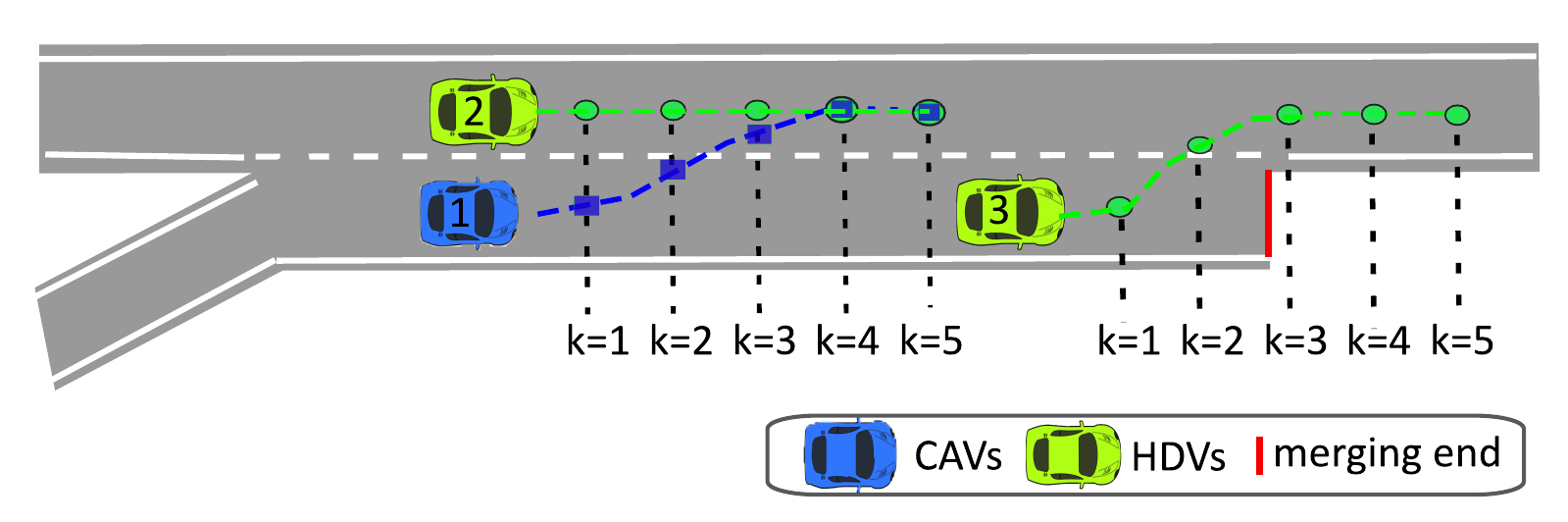}
  \caption{Illustration of trajectory conflict for $T_n=5$ steps.}
  \label{fig:conflict}
\end{figure}

\begin{figure}[!ht]
  \centering
  \includegraphics[width=0.48\textwidth]{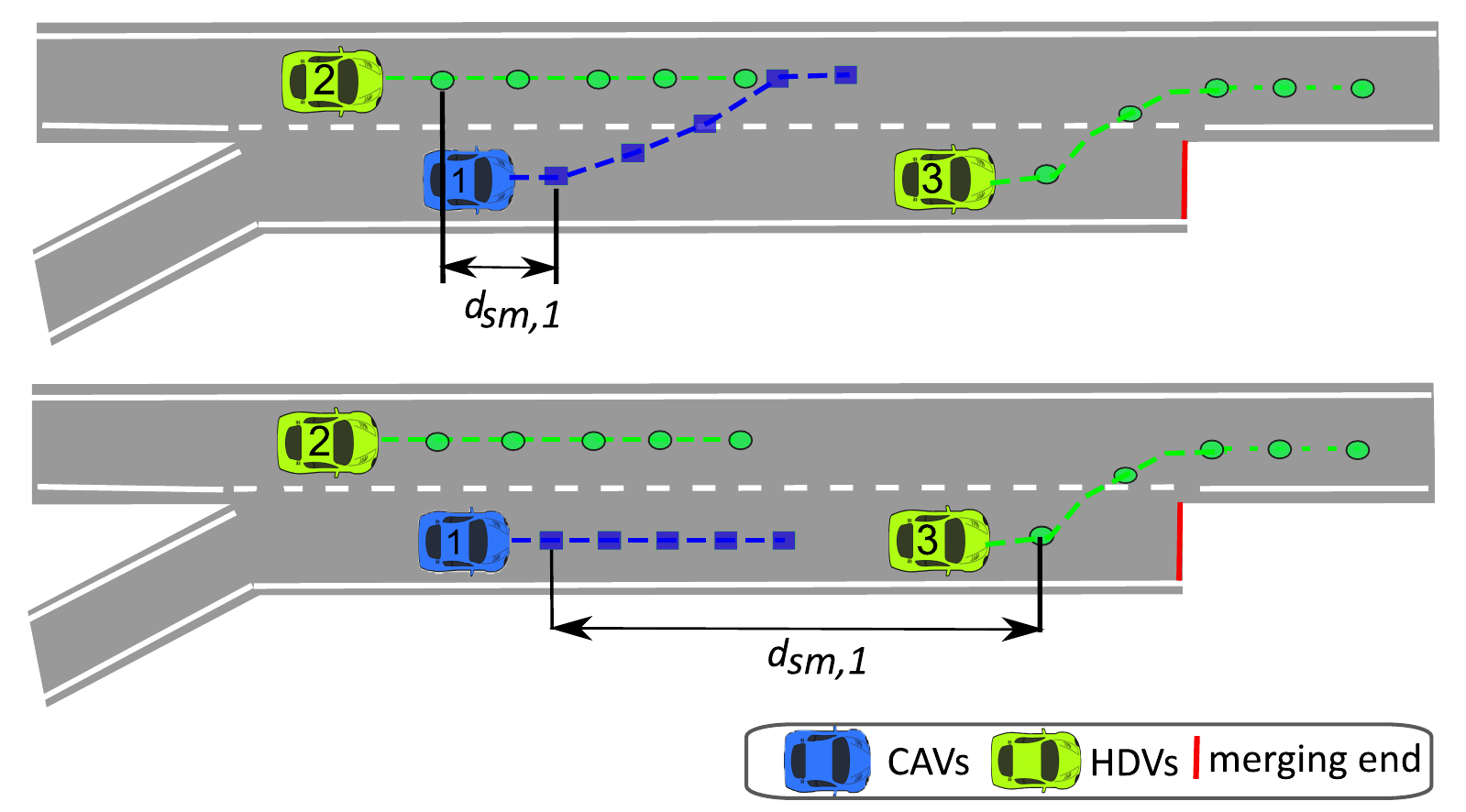}
  \caption{Illustration of safety margin definitions. Top: safety margin if vehicle 1 turns left; and Bottom: safety margin when vehicle 1 keeps straight.}
  \label{fig:safe_action}
\end{figure}

On the other hand, if the predicted trajectory of vehicle $\mathcal{P}_t[0]$ is conflicting with other considered vehicles, the exploratory action is determined as unsafe and a new ``safe'' action will be used to replace the original action. A trajectory conflict is illustrated  in Fig.~\ref{fig:conflict}, where the motion primitive of vehicle $1$ conflicts with the predicted trajectory of vehicle $2$ (HDV) at time step $4$ and $5$. The exploratory action from vehicle 1 is deemed as unsafe.
Here we assume no collisions among HDVs (i.e., rational drivers), which can (almost) be guaranteed as the IDM and MOBIL models are extremely safety-focused. Then the safety supervisor enumerate other (valid) candidate actions and pick the  best action based on the safety margin as follows:
\begin{equation}\label{eqn:safety_room}
a^{\prime}_{t} = {\arg\max}_{a_t \in \mathcal{A}_{\text{valid}}} \big(\min_{k \in T_n} d_{\text{sm}, k} \big).
\end{equation}
where $\mathcal{A}_{\text{valid}}$ is the set of valid actions at time step $t$. The safety margin $d_{\text{sm}, k}$ at the prediction time step $k$ can be obtained as follows:
\begin{itemize}
\item if the action is changing lanes, i.e., {\it turn left} or {\it turn right}, the safety margin is defined as the minimum distance to the preceding and the following vehicles on the current and target lanes. An example is shown in the top subfigure of Fig.~\ref{fig:safe_action}.
\item if the action is {\it speed up}, {\it idle}, or {\it slow down}, the safety margin is set as the minimum distance headway. An example is shown in the bottom subfigure of Fig.~\ref{fig:safe_action}.
\end{itemize}

After the action of the vehicle $\mathcal{P}_t[0]$ is decided, its trajectory can be (re-)generated. Then vehicle $\mathcal{P}_t[0]$ is removed from the list and the second highest becomes the first, i.e., $\mathcal{P}_t[i]\leftarrow \mathcal{P}_t[i+1],\,i=1,2,\cdots$. Then the same safety-check procedure discussed above is applied to the vehicle corresponding to new $P_t[0]$, except that when determining collisions, instead of using actions from the last step to generate the trajectories, the motion primitives of the  higher-priority vehicles (those who have been processed through the safety check procedure) with the safety-proved actions are used.  The procedure will continue until the priority list $\mathcal{P}_t$ is emptied. The details of the proposed priority-based safety supervisor are given in Algorithm~\ref{algo:safety_supervisor}.

\begin{remark}
The priority-based safety supervising scheme can be realized through vehicle-to-infrastructure \cite{chou2009feasibility} (V2I) communication where a central communication station near the ramp can observe the HDVs and communicate with the AVs. At each discrete time $t$, the infrastructure agent will determine the priority scores for the AVs based on the observed traffic. AVs send their exploratory actions to the infrastructure and Algorithm~1 is then performed to generate the updated, safe actions for the AVs. As Algorithm 1 is sequential and thus computationally efficient (approximately $28.13~ms$ for the safety supervisor with $T_n = 8$ to make a decision in the \textit{Hard} traffic model, see Table~\ref{tab:safety_supervisor} in Section V-C), with reasonable computation power at the infrastructure, it is expected that the algorithm can be implemented in real-time, i.e., providing updated controls within one sampling time step. For future real deployment, some further computation optimization can be performed to improve the computational efficiency, which is left for our future work.
\end{remark}

\begin{figure}
\removelatexerror
\scalebox{0.85}{
\begin{algorithm*}[H]
\SetAlFnt{\small}
    \SetKwInOut{Parameter}{Parameter}
    \SetKwInOut{Output}{Output}
\caption{Priority-based Safety Supervisor}
\label{algo:safety_supervisor}
\SetAlgoLined
\Parameter{$L, \alpha_1, \alpha_2, \alpha_3, t_h, w, T_n$.}
\Output{ $a_i, i \in \text{\larger[2]$\nu$}$.}
\vspace{0.2em}
\hrule
\vspace{0.2em}
\For{$i = 0$ to $N$}{
    compute the priority scores according to Eq.~\ref{eqn:priority_coefficient}\;
    rearrange ego vehicles to list $\mathcal{P}_t$ according to their priority scores.}
    
\For{$j= 0$ to $|\mathcal{P}_t|$}{
    obtain the highest-priority vehicle $\mathcal{P}_t[0]$\;
    find its neighboring vehicles $\mathcal{N}_{\mathcal{P_t}[0]}$\;
    predict trajectories $\zeta_{v}, v \in \mathcal{P}_t[0] \cup \mathcal{N}_{\mathcal{P_t}[0]}$ for $T_n$ time steps.
    
    \If{trajectories are overlapped}{
        replace the risky action as $a_t \leftarrow a^{\prime}_t$ according to Eq.~\ref{eqn:safety_room}\;
        replace the trajectory $\zeta_{\mathcal{P}_t[0]}$ with $\zeta^{\prime}_{\mathcal{P}_t[0]}$
    }
    remove $\mathcal{P}_t[0]$ from $\mathcal{P}_t$\;
    update $\mathcal{P}_t[i]\leftarrow \mathcal{P}_t[i+1],\,i=1,2,\cdots$.
}
\end{algorithm*}}
\end{figure}

\begin{remark}
The prediction horizon $T_n$ is an important hyper-parameter in the safety-enhancement scheme. If $T_n$ is too small, the safety supervisor is ``short-sighted'' and  can lead to no feasible solutions after a few steps. On the other hand, if $T_n$ is too large, the uncertainty of HDVs (the actual vehicle motion in the simulation has noisy perturbations from the human driver models used to predict the trajectories) are propagated and the results tend to be conservative in order to guarantee the safety in a large horizon. In our work, we use cross-validations and find that $T_n=8$ or $9$ is the best choice (see e.g., Fig.~\ref{fig:plot_reward_safety} and Table~\ref{tab:safety_supervisor}).
\end{remark}

Pseudo-code of the proposed MARL with the priority-based safety supervisor is shown in Algorithm \ref{algo:marl_algo}. 
The hyperparameters include: the (time)-discount factor $\gamma$, the learning rate $\eta$,  the total number of training epochs $M$, the epoch length $T$, and the coefficients for the loss function $\beta_1$ and $\beta_2$. In each epoch, each agent collects the state information and samples actions by applying the action masking strategy to avoid invalid actions (Lines 4--7). Then the exploratory actions from the MARL will be checked by the priority-based safety supervisor detailed in Algorithm~\ref{algo:safety_supervisor} (Line 9). If the action is unsafe, then the safety supervisor will replace the risky action with a safe action according to Eq.~\ref{eqn:safety_room}. The safe action will be taken by the agent and the corresponding experience will be collected and saved to the replay buffer (Lines 10--17). 
The parameters of the policy network are updated using the collected experience sampled from the on-policy experience buffer after the completion of each episode (Lines 20--26). The DONE signal is flagged if either the episode is completed or a collision occurs. After receiving the DONE flag, all agents are reset to their initial states to start a new epoch (Lines 28).

\begin{figure}
\removelatexerror
\scalebox{0.85}{
\begin{algorithm*}[H]
\SetAlFnt{\small}
    \SetKwInOut{Parameter}{Parameters}
    \SetKwInOut{Output}{Outputs}
\caption{MARL for AVs with Safety Supervisor}
\label{algo:marl_algo}
\SetAlgoLined
\Parameter{$\gamma, \eta, T, M, \beta_1, \beta_2$.}
\Output{$\theta$.}
\hrule
\vspace{0.2em}
{\bf initialize}  $s_0, t \leftarrow 0, \mathcal{D} \leftarrow \emptyset$;\\
\For{$j = 0$ to $M-1$}{
    	\For{$t=0$ \text{to} $T-1$}{
	          \For{$i \in \text{\larger[2]$\nu$}$}{
            	    observe $s_i$\; 
            	    update $a_{i,t} \sim \pi_{\theta_i} (\cdot | s_i)$ with action masking.
            }
            
            \For{$i \in \text{\larger[2]$\nu$}$}{
                check the actions by Algorithm~\ref{algo:safety_supervisor}\;
                \If{safe}{
                execute $a_{i,t}$\;
                update $\mathcal{D}_i \leftarrow (s_{i,t}, a_{i,t}, r_{i,t}, v_{i,t})$
                }
                
                \Else{
                update $a_{i,t} \leftarrow a^{\prime}_{i,t}$ and execute $a^{\prime}_{i,t}$ \;
                update $\mathcal{D}_i \leftarrow (s_{i,t}, a^{\prime}_{i,t}, r_{i,t}, v_{i,t})$.
                }
            }
            
        	  update $t \leftarrow t+ 1$
        	  
        	  \If{DONE}{
        	   	\For{$i \in \text{\larger[2]$\nu$}$}{
            	    update $\theta_i \leftarrow \theta_i + \eta \nabla_{\theta_i} {J(\theta_i)}$}
    	    }
    	    
    	    initialize $\mathcal{D}_i \leftarrow \emptyset, i \in \text{\larger[2]$\nu$}$\;
    	    update $j \leftarrow j+ 1$
    	}
    	update $s_0, t \leftarrow 0$
}
\end{algorithm*}}
\vspace{-15pt}
\end{figure}

\begin{figure*}[!ht]
  \centering
  \includegraphics[width=0.9\textwidth]{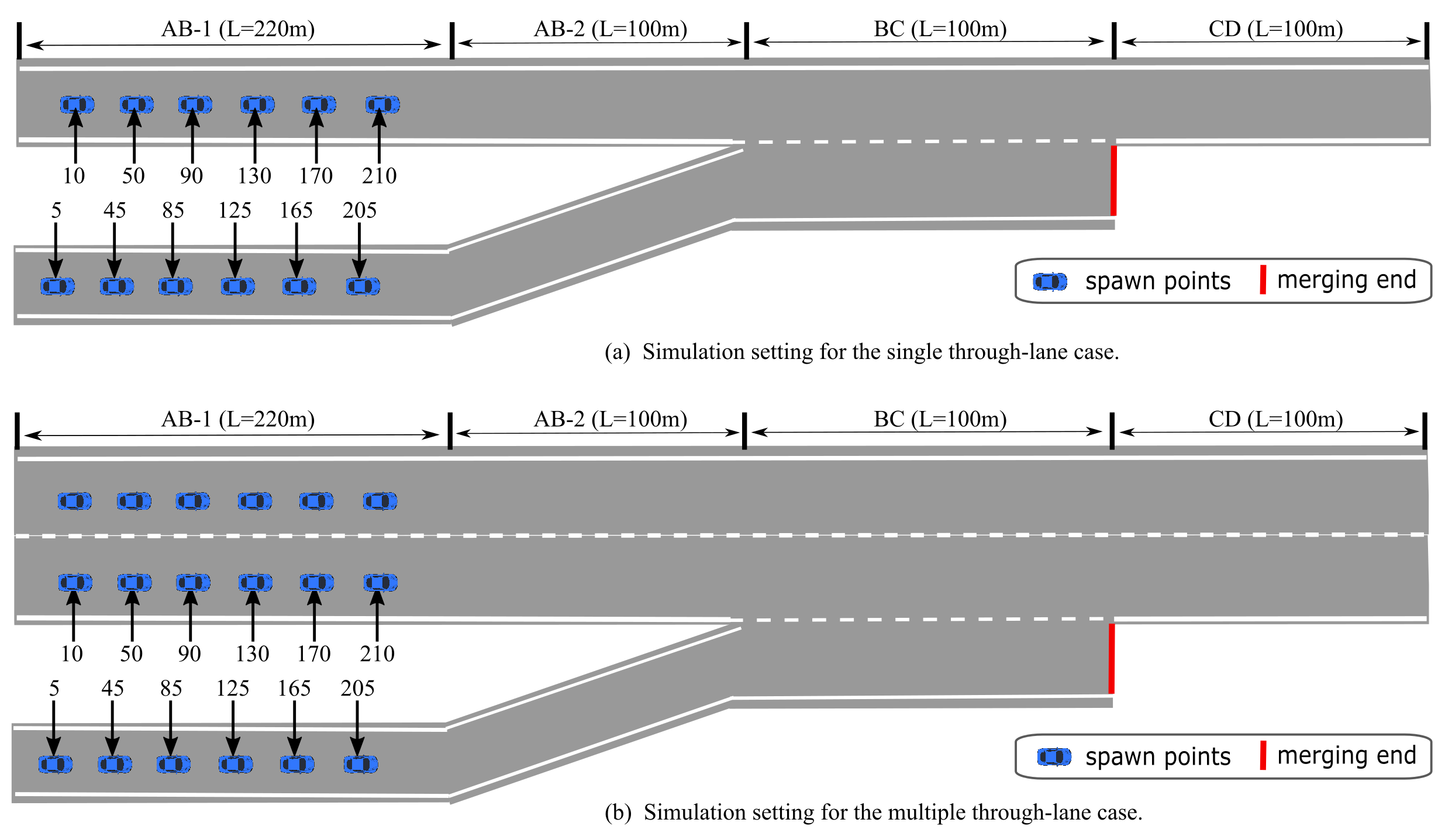}
  \caption{Simulation settings for the single through lane case (upper) and multiple through lane case (lower). ``L" represents the length of the road segments.}
  \label{fig:scenario_setting}
\end{figure*}

\section{NUMERICAL EXPERIMENTS}\label{sec:5}
In this section, we evaluate the performance of the proposed MARL algorithm in terms of training efficiency and collision rate in the on-ramp merging scenario illustrated in Fig.~\ref{fig:merging_scenario}. The length of the road is $520~m$,  where the entrance of the merge lane is at $320~m$ and the length of the merge lane is $L=100~m$. There are 12 spawn points evenly distributed on the through lane and the ramp lane from $0~m$ to $220~m$, as shown in Fig.~\ref{fig:scenario_setting}(a). The vehicles exceeding the road will be removed from displaying while the kinematics are still updated.
Specifically, we consider three levels of traffic densities with different number of initial vehicles defined as:
\begin{itemize}
    \item \textit{Easy mode}: 1-3 AVs and 1-3 HDVs.
    \item \textit{Medium mode}: 2-4 AVs + 2-4 HDVs.
    \item \textit{Hard mode}: 4-6 AVs + 3-5 HDVs.
\end{itemize}

In each training episode, a different number of HDVs and AVs will randomly appear at the spawn points with a random position noise (uniformly distributed in [-1.5m, 1.5m]) added to each initial spawn position. The initial speed is randomly chosen between 25 to 27 $m/s$. The vehicle control sampling frequency is 5 Hz, i.e., AVs take an action every 0.2 seconds. A 5\% random noise is added to the predicted acceleration and steering angle for HDVs.
We train all MARL algorithms over 2 million steps with 3 different random seeds while the same random seeds are shared among the agents, which is around 20,000 episodes with episode horizon $T = 100$ steps. We evaluate the algorithm over 3 episodes every 200 training episodes. We set $\gamma=0.99$ and the learning rate $\eta = 5e^{-4}$; The coefficients $w_c$, $w_s$, $w_h$, and $w_m$ for the reward function are set as 200, 1, 4, and 4, respectively. The priority coefficients $\alpha_1$, $\alpha_2$ and $\alpha_3$ are equally set as 1. The weighting coefficients $\beta_1$ and $\beta_2$ for the loss function are chosen as 1 and 0.01, respectively. Here we call the MARL algorithm, without the safety supervisor, proposed in Section~III as the baseline method. 

The simulation environment is modified from the gym-based highway-env simulator~\cite{highway-env} and is open-sourced\footnote{See \url{https://github.com/DongChen06/MARL_CAVs}}. We use the default parameters of the IDM and MOBIL models which can be found as in the highway-env simulator~\cite{highway-env}. The experiments have been performed in a Ubuntu 18.04 server with AMD 9820X processor and 64 GB memory. The video demo of the training process can be found at the site\footnote{See \url{https://drive.google.com/drive/folders/1437My4sDoyPFsUjrThmlu1oJjTkTkvJ7?usp=sharing}}.

\subsection{Reward Function Designs}\label{sec:5-1}
In this subsection, we will first evaluate the performance of the proposed MARL framework under different reward function designs, local (baseline) v.s. global rewards (baseline with global reward). Then the impact of the safety penalty weight $w_c$ in the reward function (Eq.~\ref{eqn:reward_function}) will be evaluated.

\begin{figure*}[!ht]
  \centering
  \includegraphics[width=0.95\textwidth]{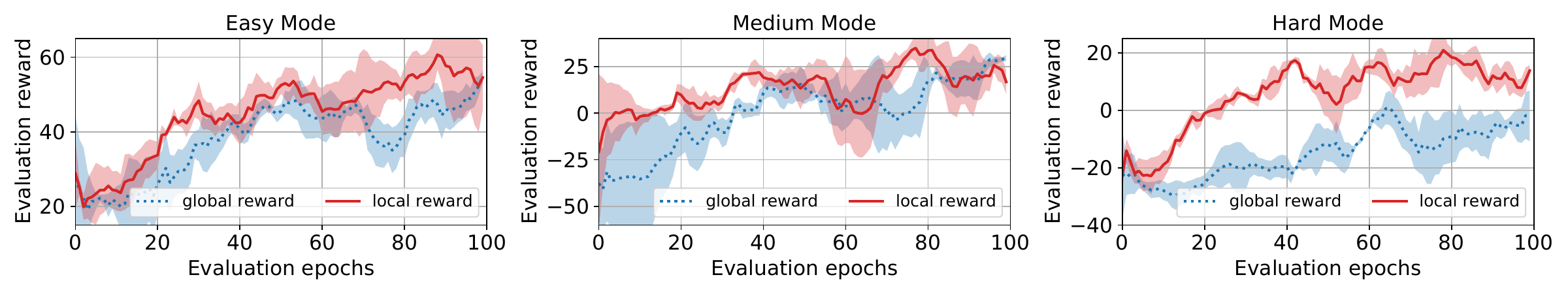}
  \caption{Evaluation curves during training with different reward functions for different traffic levels. The shaded region denotes the standard deviation over 3 random seeds. The curves are smoothed over the last 9 evaluation epochs.}
  \label{fig:plot_reward_fn}
\end{figure*}

We investigate the proposed local reward function by comparing it with the global reward design used in \cite{dong2020drl, kaushik2018parameter}, where the reward of the $i$th agent at time step $t$ is the averaged global reward $r_{i, t} = \frac{1}{N} \sum_{j=1}^{N} r_{j,t}$.
Fig.~\ref{fig:plot_reward_fn} shows the evaluation performance comparison between the proposed local reward design and the global reward design. As expected, the proposed local reward outperforms the global reward design in terms of higher evaluation rewards as well as faster convergence speed across all three traffic scenarios. In the {\it Easy} and {\it Medium} modes, the global reward design performs well and achieves a reasonable reward due to the small number of AVs, while it fails the control tasks in the {\it Hard} mode, with an evaluation reward less than 0, as it suffers from the credit assignment issues \cite{sutton2018reinforcement} and the fact that the assigned average global rewards have less correlation with individual agent’s actions as the number of agents increases.

Table~\ref{tab:wc} shows the testing performance with different $w_c$ values in the \textit{Medium} traffic mode, while other weighting coefficients are kept unchanged in Eq.~\ref{eqn:reward_function}. It can be seen that there are no collisions when a large enough $w_c$ is selected (e.g., $w_c\geq100$), and the average traffic speed decreases as $w_c$ further increases. This is because the CAVs will behave conservatively if we emphasize too much on the safety.  In the following experiments, we set $w_c$ to 200 since it represents a reasonable trade-off between safety margin and traffic efficiency.

\begin{table*}[!ht]
\renewcommand{\arraystretch}{1.4}
\centering
\caption{Performance with different $w_c$'s in terms of collision rate and average speed in the \textit{Medium} traffic mode. Note that other weighting factors are kept unchanged.}
\label{tab:wc}
\begin{tabular}{c|c|c|c|c|c}
\hline
        & $w_c = 10$ & $w_c=100$ & $w_c = 200$ (we chose) & $w_c = 1000$ & $w_c = 10000$ \\ \hline
Collision rate & 0.1       & 0        & 0          & 0           & 0            \\ \hline
avg. speed     & 24.77     & 24.09    & 24.08      & 23.69       & 23.62        \\ \hline
\end{tabular}
\end{table*}

\subsection{Curriculum Learning}
In this subsection, we adopt curriculum learning \cite{kaushik2018overtaking} to speed up the learning and improve the performance for the {\it Hard} mode. Specifically, instead of learning the {\it Hard} mode directly,  we build upon the trained model from the easier modes (i.e., {\it easy} and {\it medium}) and train the models to achieve higher efficiency. Curriculum learning is especially preferable to safety-critical tasks (e.g., autonomous driving) as starting from a decent model can greatly reduce the number of ``blind'' explorations that can be risky.

\begin{figure}[!ht]
  \centering
  \includegraphics[width=0.43\textwidth]{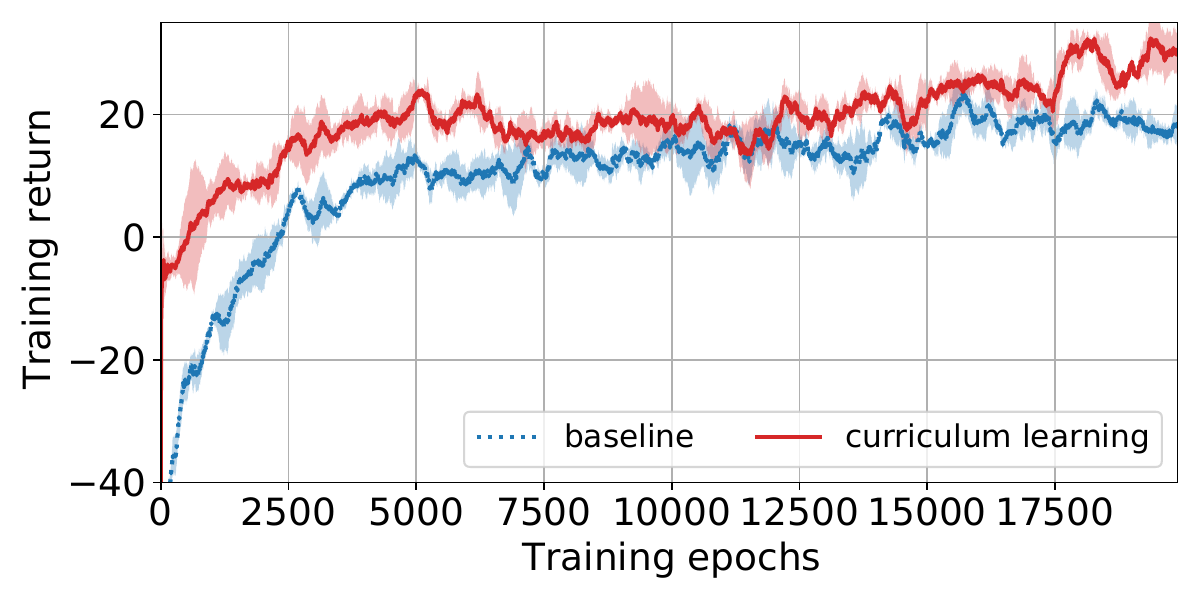}
  \caption{Training curves with and without curriculum learning for {\it Hard} traffic mode.}
  \label{fig:plot_curr}
\end{figure}

\begin{figure}[!ht]
  \centering
  \includegraphics[width=0.43\textwidth]{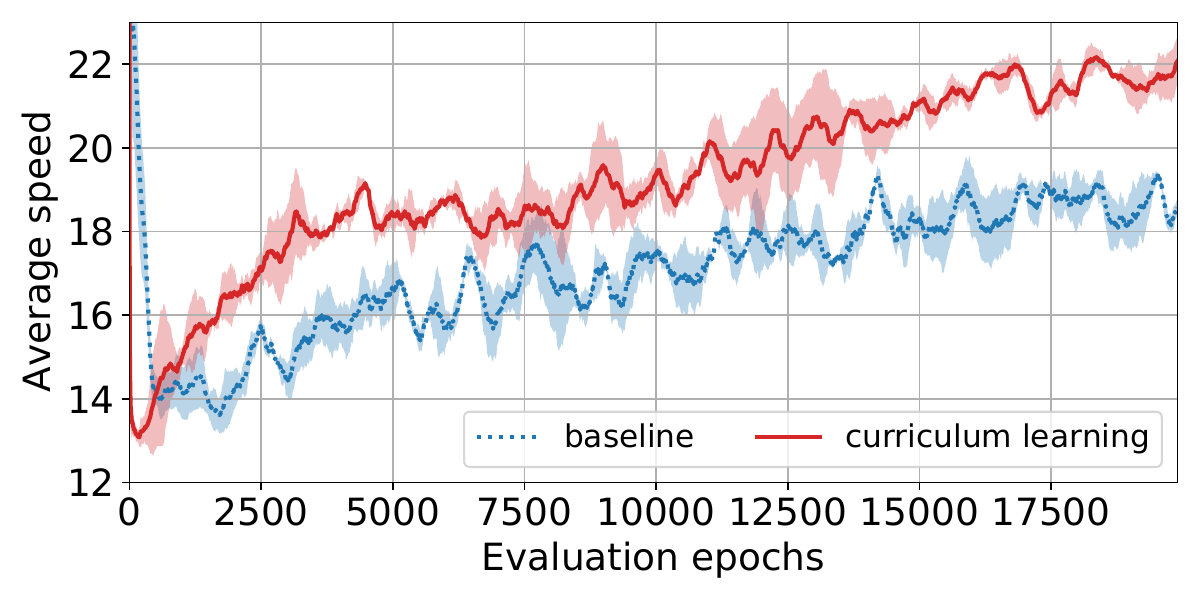}
  \caption{Average speed during training with and without curriculum learning for {\it Hard} traffic mode.}
  \label{fig:plot_curr_speed}
  \vspace{-10pt}
\end{figure}

Fig.~\ref{fig:plot_curr} shows training performance comparison between the baseline method (i.e., starting from scratch) and curriculum learning (baseline + curriculum learning) for the {\it Hard} traffic mode. It is obvious that learning based on the trained model from easier tasks greatly expedites the speed of convergence and improves the final model performance. The average speed during the training, as shown in Fig.~\ref{fig:plot_curr_speed}, indicates that the curriculum learning strategy also improves the average vehicle speed up to $22~m/s$ compared to baseline method at $18~m/s$, thus achieving high traffic efficiency. Therefore, we apply the curriculum learning in the following experiments for {\it Hard} traffic modes.

\subsection{Performance of the Priority-based Safety Supervisor}\label{sec:V-3}
\begin{figure*}[!ht]
  \centering
  \includegraphics[width=0.98\textwidth]{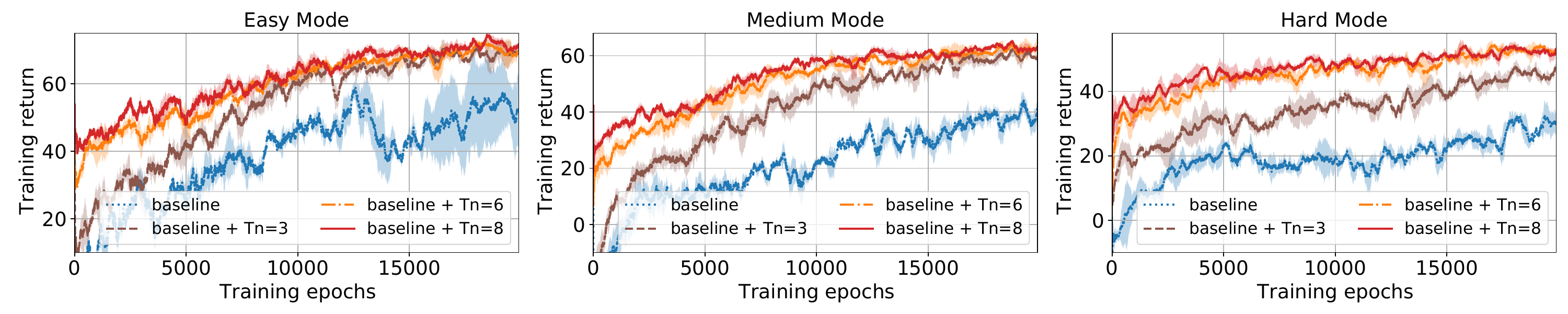}
  \caption{Training curves for the n-step priority-based safety supervisor.}
  \label{fig:plot_reward_safety}
\end{figure*}

In this subsection, we evaluate the effectiveness of the proposed priority-based safety supervisor. 
As shown in Fig.~\ref{fig:plot_reward_safety}, the proposed priority-based safety supervisor method has a better sample efficiency, evidenced by faster converge speed in all three traffic densities. In addition, the proposed priority-based safety supervisor method achieves higher evaluation reward even in the \textit{Hard} traffic mode.
This is because most unsafe actions are replaced with safe ones by the safety supervisor, especially in the earlier exploration phase, which avoids early terminations and thus improves learning efficiency. 

\begin{figure*}[!ht]
  \centering
  \includegraphics[width=0.98\textwidth]{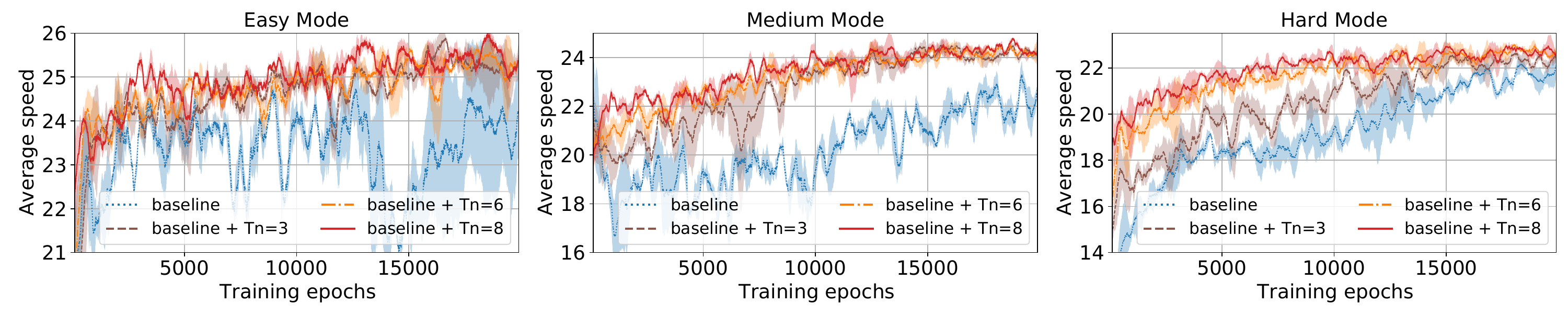}
  \caption{Average speed during training for the n-step priority-based safety supervisor.}
  \label{fig:plot_reward_safety_speed}
\end{figure*}

Fig.~\ref{fig:plot_reward_safety_speed} shows the average vehicle speed during the training, which is an indication of traffic throughput. It is clear that the algorithms with the safety supervisor maintain higher training speed than the baseline method (i.e., without safety supervision). This shows that the proposed safety supervisor is not only beneficial for training but also leads to better traffic efficiency. It can also be seen that  vehicle speeds are slower as the traffic density increases ($26~m/s$, $24~m/s$ and $22~m/s$ for \textit{Easy}, \textit{Medium} and \textit{Hard} traffic densities, respectively), which is reasonable as the interactions are more frequent in a dense traffic and lower speed is safer to avoid collisions.

After training, MARL algorithms for each traffic density are tested over 3 random seeds for 30 epochs and the average collision rates, vehicle speeds, and inference time of the safety supervisor are shown in Table~\ref{tab:safety_supervisor}. We can see that with the safety supervisor ($T_n >= 7$), the MARL can run without collisions in all traffic modes, while the baseline method has a collision rate of $0.07$ and $0.16$ for \textit{Medium} and \textit{Hard} traffic densities, respectively. It is clear that with only a short prediction horizon, e.g., $T_n = 3~or~6$, the MARL is still failed under challenging cases. For example, the agents still have $0.03$ collision rate in the \textit{Hard} traffic mode when choosing $T_n = 6$.  The reason is that if $T_n$ is too small, the safety supervisor is ``short-sighted'' and can lead to no feasible solutions after only a few steps. However, if we further increase $T_n$ (e.g., $T_n = 10, 12, 14$), the collision rate may increase, as discussed in Remark 2 in Section IV-B, due to the uncertainty propagation in a longer time window. Also, the average speed will decrease since the CAVs need to behave carefully to guarantee the safety in a large time horizon.
With a reasonable $T_n$ (e.g., 7, 8), the average speed indicates that the safety supervisor leads to higher traffic efficiency. In all traffic modes, safety supervisor always leads to higher average speed while lower collision rate. For example, the best average speed for the \textit{Easy} traffic mode is achieved by baseline + $T_n=8$ ($27.72~m/s$) compared to the baseline method ($23.52~m/s$).
As expected, the
inference time increases as the prediction horizon increases. With reasonable computation power at the infrastructure, it is expected that the algorithm can be implemented in real-time.

\begin{table*}[!ht]
\renewcommand{\arraystretch}{1.4}
\centering
\caption{Testing performance comparison of collision rate, average speed (m/s), and inference time (ms) between n-step safety supervisor based on the baseline (bs) method.}
\label{tab:safety_supervisor}
\resizebox{0.98 \textwidth}{!}{%
\begin{tabular}{c|c|c|c|c|c|c|c|c|c|c}
\hline
Scenarios                    & Metrics          & bs        & bs + $T_n=3$ & bs + $T_n=6$  & bs + $T_n=7$ & bs + $T_n=8$        & bs + $T_n=9$ & bs + $T_n=10$ & bs + $T_n=12$        & bs + $T_n=14$ \\ \hline
\multirow{2}{*}{Easy Mode}   & collision rate & 0  & 0 & 0        & 0       & 0     & 0   & 0       & 0     & 0     \\ \cline{2-11} 
                             & avg. speed   & 23.53 & 25.12  & 25.38       & 25.27   & 27.72 & 27.50  & 25.89    & 25.82     & 25.74 \\ \cline{2-11} 
                             & infrn time     & - & 4.90 & 7.62 & 8.30   & 8.93      & 10.07 & 11.08   & 12.62     & 14.71
                            \\ \hline
\multirow{2}{*}{Medium Mode} & collision rate  & 0.07  & 0.03 & 0.01        & 0    & 0  & 0  & 0       & 0     & 0.01  \\ \cline{2-11} 
                             & avg. speed     & 20.30  & 24.22  & 24.61     & 24.13   & 24.08 & 24.19  & 23.74       & 24.35     & 24.13 \\ \cline{2-11} 
                             & infrn time     & - & 14.75 & 14.64 & 16.78   & 17.55          & 19.40 & 21.29   & 23.22     & 26.89
                            \\ \hline
\multirow{2}{*}{Hard Mode}   & collision rate  & 0.16   & 0.14   & 0.03     & 0    & 0  & 0  & 0.03       & 0.05 & 0.05  \\ \cline{2-11} 
                             & avg. speed    & 21.71 & 22.52 & 22.56 & 22.58   & 22.73          & 23.01 & 22.52       & 21.31     & 21.83  \\ \cline{2-11} 
                             & infrn time    & - & 14.75 & 23.13 & 25.69   & 28.13          & 31.45 & 35.93   & 39.43     & 50.13 \\ \hline
\end{tabular}
}
\vspace{-10pt}
\end{table*}

\begin{figure*}[!ht]
  \centering
  \includegraphics[width=0.98\textwidth]{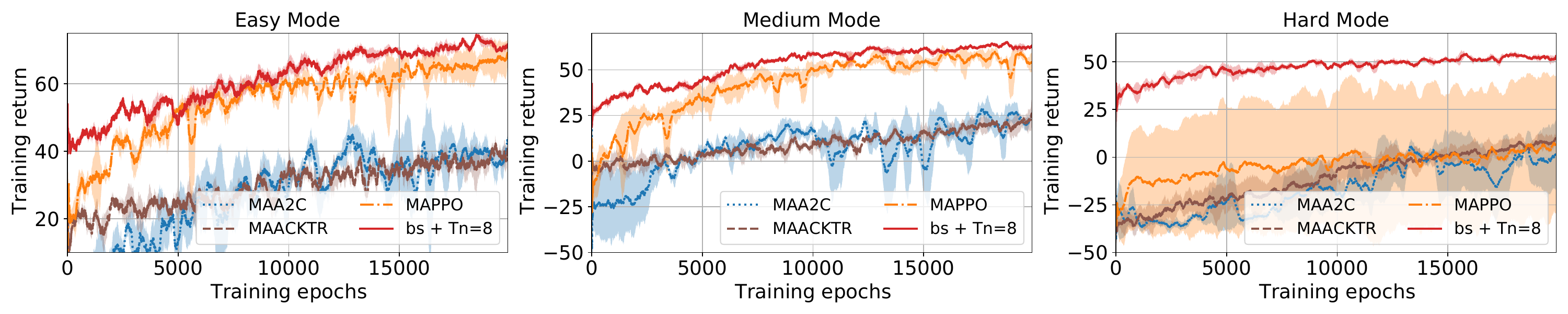}
  \caption{Training curves comparison between the proposed MARL policy (baseline (bs) + $T_n=8$) and 3 state-of-the-art MARL benchmarks.}
  \label{fig:plot_reward_safety_benchmark}
\end{figure*}

\begin{figure*}[!ht]
  \centering
  \includegraphics[width=0.98\textwidth]{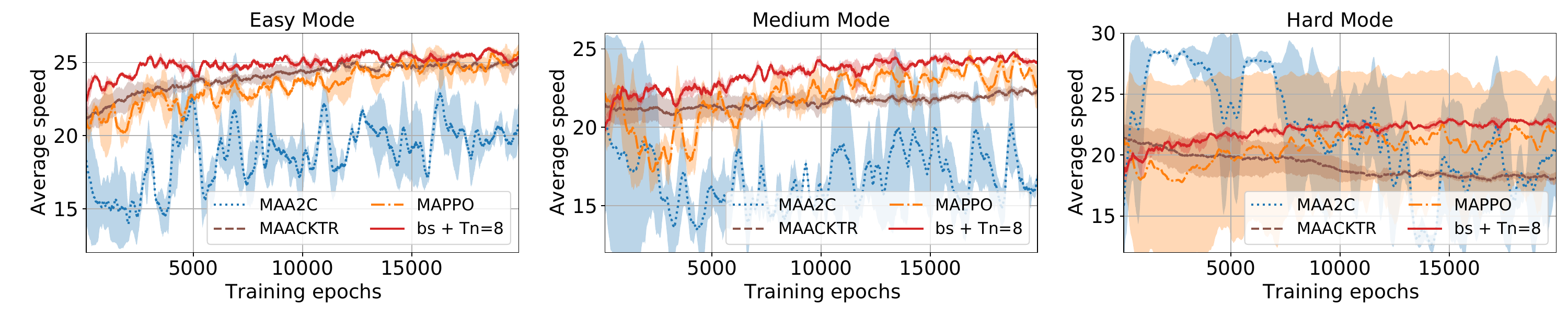}
  \caption{Average speed comparison between the proposed MARL policy (baseline (bs) + $T_n=8$) and 3 state-of-the-art MARL benchmarks.}
  \label{fig:plot_reward_safety_speed_benchmark}
\end{figure*}

\subsection{Comparison with State-of-the-art Benchmarks}

\begin{table*}[!ht]
\renewcommand{\arraystretch}{1.4}
\centering
\caption{Testing performance comparison of collision rate and average speed between the proposed method and 4 state-of-the-art benchmarks.}
\label{tab:collision_benchmarks}
\begin{tabular}{c|c|c|c|c|c|c}
\hline
Scenarios                    & Metrics    & MPC      & MAA2C          & MAACKTR & MAPPO & baseline + $T_n=8$       \\ \hline
\multirow{2}{*}{Easy Mode}   & collision rate  & 0.03  & 0.02           & 0.08    & 0  & 0     \\ \cline{2-7} 
                             & avg. speed  [m/s] & 22.05 & 21.00          & 24.71   & 25.70 & 25.72 \\ \hline
\multirow{2}{*}{Medium Mode} & collision rate & 0.03 & 0.08           & 0.12    & 0.02  & 0  \\ \cline{2-7} 
                             & avg. speed  [m/s] & 19.67    & 19.33          & 21.94   & 24.00 & 24.08 \\ \hline
\multirow{2}{*}{Hard Mode}   & collision rate & 0.40  & 0.52           & 0.18    & 0.34  & 0  \\ \cline{2-7} 
                             & avg. speed  [m/s] & 21.02   &19.68 & 18.19   & 22.41 & 22.73          \\ \hline
\end{tabular}
\vspace{-10pt}
\end{table*}

In this subsection, we compare the proposed method with several state-of-the-art MARL benchmarks, MAA2C, MAPPO and MAACKTR, as mentioned in Section~\ref{sec:2}, as well as an improved model predictive control (MPC) method \cite{cao2013two,cao2015cooperative}. The dynamics, cost function and constraint ingredients of the improved MPC approach are elaborated in the Appendix.
All the MARL benchmarks are implemented by sharing parameters among agents to deal with dynamic numbers of agents, and using the global reward and the discrete action space. We adopt the 8-step safety supervisor (baseline + $T_n=8$) in all traffic levels as it takes a good trade-off between collision rate and prediction efficiency as shown in Section~\ref{sec:V-3}.  

Fig.~\ref{fig:plot_reward_safety_benchmark} shows the evaluation results during training for all the MARL algorithms. The proposed method (baseline + $T_n=8$) consistently outperforms the benchmarks in all traffic levels. The proposed method shows even greater advantage in terms of sample efficiency and training performance in the {\it Hard} mode over the benchmarks.
Fig.~\ref{fig:plot_reward_safety_speed_benchmark} shows the proposed method has relatively higher average training speeds which lead to high training efficiency. Note that it is not wise to have high vehicle speed as in MAPPO methods in dense traffic, which will lead to very high collision rates as shown in Table~\ref{tab:collision_benchmarks}.

After training, all algorithms for each traffic density are tested over 3 random seeds for 30 epochs and the average collision rates and vehicle speeds are shown in Table~\ref{tab:collision_benchmarks}. The testing results also show that the proposed method can run without collisions and achieve higher efficiency than other benchmark algorithms.
Also, it is noteworthy that, due to the discrepancy between the exact dynamics used in the highway simulator environment and our model used in MPC, along with the uncertainties injected in the simulator, the MPC can still lead to collisions (0.03, 0.03 and 0.40 collision rate in the \textit{Easy},  \textit{Medium}, and \textit{Hard} traffic modes, respectively).
Another observation is on the impact of system complexity on the reliability/performance of model-based methods: when the number of vehicles grows, the merging problem becomes more difficult due to increased model mismatch (system state dimension is higher with more vehicles), and MPC finds it harder to obtain a collision-free policy for on-ramp merging of CAVs. This shows an advantage of model-free approaches that do not rely on explicit models. 
Another consideration for the model-based MPC implementation is the requirement for powerful computational resources to support the extensive online computations. This is especially significant for the on-ramp merging problem that involves nonlinear dynamics, and a nonlinear program must be solved at each time step in order to calculate the control input, requiring significant onboard computation power.
In contrast, our RL-based approach requires much less computational capability.

\begin{figure}[!ht]
  \centering
  \includegraphics[width=0.48\textwidth]{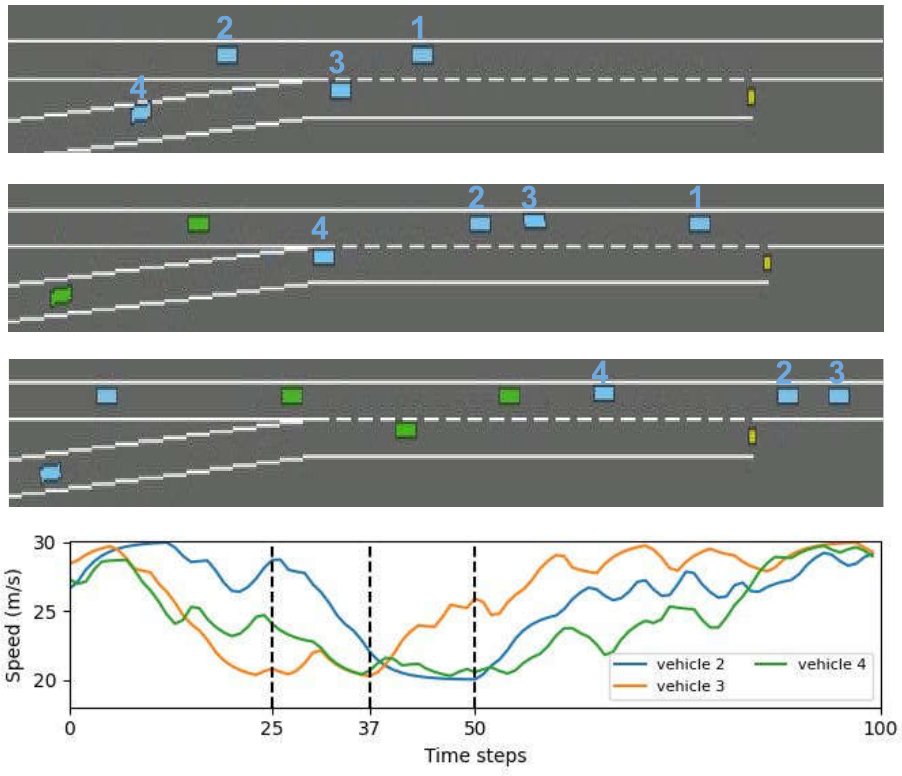}
  \caption{Frames show the learned policy. Below figure shows the corresponding speed of the AVs.}
  \label{fig:plot_learned_policy}
  \vspace{-15pt}
\end{figure}

\subsection{Policy Interpretation}
In this subsection, we attempt to interpret the learned AV behaviors. As an example, Fig.~\ref{fig:plot_learned_policy} shows the snapshots at time steps $25$, $37$, and $50$, as well as the speeds of agents 2-4. It can be observed that at the time step $25$, vehicle $2$ starts to slow down and makes space for vehicle $3$ to merge. Vehicle $3$ accelerates to merge while keeping an adequate distance headway with vehicle $1$. Then vehicle $3$ successfully merges into the through lane and starts to speed up at time step $37$. At the same time, vehicle $2$ still keeps a low speed to keep a safe headway distance to vehicle $3$. At time $50$, vehicle $2$ speeds up while keeping a lower speed than vehicle $3$ to maintain a safe distance headway. Similar patterns are also observed in vehicle $4$.

\subsection{Multiple Through-lane Case}
In this subsection, we demonstrate the proposed approach in a more challenging multiple through-lane case  illustrated in Fig.~\ref{fig:scenario_setting}(b) where vehicles are allowed to change lanes in the through lanes. As shown in Section~III-A, we formulated the on-ramp merging as a partially observable Markov decision process (POMDP) $\mathcal{M_G}$, which can be described by the following tuple $(\{\mathcal{A}_i, \mathcal{S}_i, \mathcal{R}_i\}_{i\subseteq \nu}, \mathcal{T})$. In this setting, the action state $\mathcal{A}$ is extended to the multiple through-lane case without changes, while the state space is slightly modified to accommodate for more surrounding vehicles. Specifically, the observation space (number of observable neighboring vehicles) is determined by the parameter $N_{\mathcal{N}_i}$. For the multiple through-lane case, we choose a larger $N_{\mathcal{N}_i} = 8$ ($N_{\mathcal{N}_i} = 5$ in the single through-lane case). For the priority-based safety supervisor, we also extend it to the multi-lane case without any changes.

\begin{figure}[!ht]
  \centering
  \includegraphics[width=0.49\textwidth]{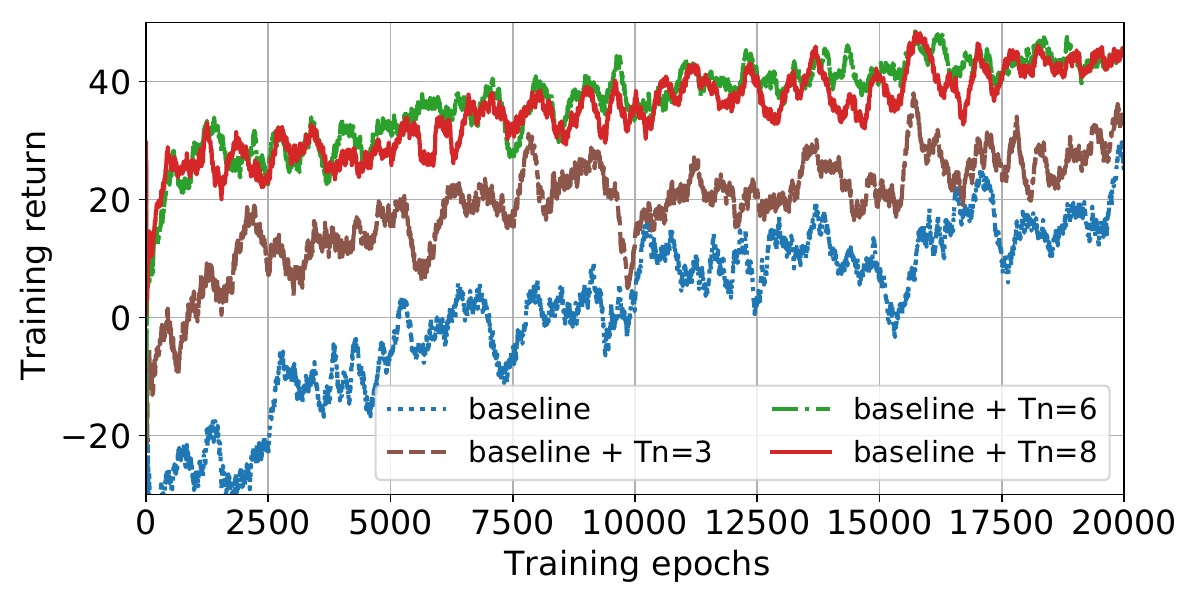}
  \caption{Training curves for the n-step priority-based safety supervisor for the multiple through-lane case.}
  \label{fig:training_multi_lane}
  \vspace{-10pt}
\end{figure}

For the reward function, we first tried to use the original reward design but we then found that the ego vehicles often conducting unnecessary and frequent lane changes which lead to unsafe driving (a demo on frequent lane changes at \url{https://drive.google.com/file/d/1dO8xPCwLXVRgQFM_xwqscRazoId5ksf4/view?usp=sharing}). Therefore, we added one more metric $r_l$ and use the following 
revised reward function:
\begin{equation}\label{eqn:reward_function}
r_{i,t} = w_c r_c + w_s r_s + w_h r_h + w_m r_m + w_l r_l,
\end{equation}
where $w_c$, $w_s$, $w_h$, $w_m$ and $w_l$ are positive weighting scalars corresponding to collision evaluation $r_c$, stable-speed evaluation $r_s$, headway time evaluation $r_h$, merging cost evaluation $r_m$, and lane-changing evaluation $r_l$, respectively. The goal of the  added lane-changing evaluation $r_l$ is to penalize unnecessary and frequent lane changes to avoid oscillatory driving following the designs in \cite{saxena2020driving}. Here, $r_l$ is defined as:
\begin{equation}
    r_l = 
    \begin{cases}
          -1, & \text{if change lanes}; \\
          0, &  \text{otherwise}.
    \end{cases}
\end{equation}
\\
To further demonstrate the flexibility and effectiveness of the proposed MARL framework, we implemented the aforementioned multiple through-lane cases in the highway environment. Fig.~\ref{fig:training_multi_lane} shows that our approach can be easily extended to the multiple through-lane case and achieves good performance. Demo video and code for the multiple through-lane scenario can be found at \url{https://github.com/DongChen06/MARL_CAVs/tree/multi-lane}. For more comprehensive testing and comparison with other benchmarks, we will explore them in the future work.

\section{Conclusions and Discussions}\label{sec:6}
In this paper, we formulated the problem of on-ramp merging in a mixed-traffic as an on-policy MARL, and we developed an efficient MARL algorithm featuring  action masking, local reward design, curriculum learning, and parameter sharing. A novel priority-based safety supervisor was also developed to enhance safety, improve learning efficiency, and increase traffic throughput. Comprehensive experiments were conducted to compare with several state-of-the-art algorithms, which showed that the proposed approach consistently outperformed the benchmark approaches in terms of training efficiency and collision rate. 

In the future work, we plan to investigate how to fill the gap between simulations and real-world implementations. The initial exploration in general RL training may lead to undesired system behavior and sometimes can even cause crashes in real-world deployment. For such systems, the safety during exploration can be enhanced by exploiting the dynamic information of the system to limit the exploration actions within an admissible range, see e.g., our previous work \cite{chen2020autonomous} as well as others \cite{alshiekh2018safe, garcia2015comprehensive} for safe RL algorithms. Before real-world implementations, the policy network is necessary to be trained in high-fidelity simulations  until sufficiently good performance is achieved, which can also significantly reduce the number of risky explorations. Also, it needs to pass various tests before field deployment for safety and robustness. Once the policy is deployed on the ego vehicles, periodical updates and maintenance should be conducted to improve the model performance in unseen scenarios. Interested readers are referred to an extensive survey on sim-to-real paradigms for reinforcement learning \cite{zhao2020sim}. Therefore, we will develop a more realistic simulation environment by incorporating data from real-world traffic systems to better fill the sim2real gap.

On the other hand, in this paper, the built-in lane-changing behavior of HDVs in the Highway simulator \cite{highway-env} is governed by the Minimizing Overall Braking Induced by Lane change (MOBIL) model \cite{kesting2007general}, which is not suitable for highway on-ramp scenarios since the HDVs are not aware of the merging end in the MOBIL model. As a result, sometimes, the HDVs may incur unexpected delays on the ramp, see the video demo (blue: CAVs, green: HDVs) at: \url{https://drive.google.com/file/d/1mWFgZrsX4bbs0M-D-fmLQ_oK5V7OwjWU/view?usp=sharing}. Therefore, in the future work, we will also incorporate more comprehensive human driver models so that HDV motions can be more accurately predicted and the overall traffic statistics (e.g., throughput, queue and delay on the ramp) can be studied.

\appendix
In this paper, we compared our proposed MARL approach with a model-based control strategy, i.e. Model Predictive Control (MPC).
In this regard,
we mainly adopted the MPC problem formulation proposed in \cite{cao2013two,cao2015cooperative} for the highway on-ramp merging problem. To capture the vehicle interactions and dynamics in the highway simulator environment, we improved the MPC implementation in \cite{cao2013two,cao2015cooperative} following the specifications:
\begin{enumerate}
    \item The following elements are employed in our problem formulation to simulate vehicle dynamics.
     \begin{itemize}
         \item Kinematic Bicycle Model \cite{polack2017kinematic} is used for vehicle kinematics instead of the point mass model used in \cite{cao2013two,cao2015cooperative}.
         \item For longitudinal behavior of HDVs, the acceleration of the HDVs is given by the Intelligent Driver Model (IDM) from \cite{treiber2000congested}. For lateral behavior, the discrete lane change decisions of HDVs are modeled by the Minimizing Overall Braking Induced by Lane change (MOBIL) model
        \cite{kesting2007general}.
        \item The CAVs control inputs are the steering angle and the acceleration. This could be translated to a high-level decisions (i.e. faster, slower, idle, left-lane) similar to the action space used in the MARL framework.
     \end{itemize}
    \item The cost function is made up of the following components:
        \begin{itemize}
        \item Collision avoidance term (i.e., Eq. 8 in the reference \cite{cao2015cooperative}) in the cost function is adopted.
        \item To bound the available moving area for vehicles, a cost term similar to Eq. 13 and Eq. 14 in \cite{cao2015cooperative} is incorporated in the cost function.
        \item The nominal path for the merging CAVs is defined based on the geometry of the on-ramp lane and the main lane, similar to Eq. 3 in \cite{cao2015cooperative}.
        \item To make vehicles run as closely as possible at their desired speed, a term similar to Eq. 15 in \cite{cao2015cooperative} is included in the cost function.
     \end{itemize}
     \item Hard constraints on the acceleration and steering angle are considered.
\end{enumerate}
A Nonlinear MPC problem is formulated and implemented following the above descriptions, with the results shown in Table~\ref{tab:collision_benchmarks}. The detailed implementation of the MPC approach is open-source and can be accessed at: \url{https://github.com/DongChen06/MARL_CAVs/tree/MPC_Merging}.

\bibliography{ref}

\begin{thebibliography}{10}
\providecommand{\url}[1]{#1}
\csname url@samestyle\endcsname
\providecommand{\newblock}{\relax}
\providecommand{\bibinfo}[2]{#2}
\providecommand{\BIBentrySTDinterwordspacing}{\spaceskip=0pt\relax}
\providecommand{\BIBentryALTinterwordstretchfactor}{4}
\providecommand{\BIBentryALTinterwordspacing}{\spaceskip=\fontdimen2\font plus
\BIBentryALTinterwordstretchfactor\fontdimen3\font minus
  \fontdimen4\font\relax}
\providecommand{\BIBforeignlanguage}[2]{{%
\expandafter\ifx\csname l@#1\endcsname\relax
\typeout{** WARNING: IEEEtran.bst: No hyphenation pattern has been}%
\typeout{** loaded for the language `#1'. Using the pattern for}%
\typeout{** the default language instead.}%
\else
\language=\csname l@#1\endcsname
\fi
#2}}
\providecommand{\BIBdecl}{\relax}
\BIBdecl

\bibitem{autopilot}
``Future of driving,'' \url{https://www.tesla.com/autopilot }, accessed:
  2021-03-31.

\bibitem{apollo}
``Apollo open platform,'' \url{https://apollo.auto/developer.html}, accessed:
  2021-03-31.

\bibitem{dixit2016autonomous}
V.~V. Dixit, S.~Chand, and D.~J. Nair, ``Autonomous vehicles: disengagements,
  accidents and reaction times,'' \emph{PLoS one}, vol.~11, no.~12, p.
  e0168054, 2016.

\bibitem{favaro2017examining}
F.~M. Favar{\`o}, N.~Nader, S.~O. Eurich, M.~Tripp, and N.~Varadaraju,
  ``Examining accident reports involving autonomous vehicles in california,''
  \emph{PLoS one}, vol.~12, no.~9, p. e0184952, 2017.

\bibitem{ni2005simplified}
D.~Ni and J.~D. Leonard~II, ``A simplified kinematic wave model at a merge
  bottleneck,'' \emph{Applied mathematical modelling}, vol.~29, no.~11, pp.
  1054--1072, 2005.

\bibitem{leclercq2011capacity}
L.~Leclercq, J.~A. Laval, and N.~Chiabaut, ``Capacity drops at merges: An
  endogenous model,'' \emph{Procedia-Social and Behavioral Sciences}, vol.~17,
  pp. 12--26, 2011.

\bibitem{bouton2019cooperation}
M.~Bouton, A.~Nakhaei, K.~Fujimura, and M.~J. Kochenderfer, ``Cooperation-aware
  reinforcement learning for merging in dense traffic,'' in \emph{2019 IEEE
  Intelligent Transportation Systems Conference (ITSC)}.\hskip 1em plus 0.5em
  minus 0.4em\relax IEEE, 2019, pp. 3441--3447.

\bibitem{rios2016survey}
J.~Rios-Torres and A.~A. Malikopoulos, ``A survey on the coordination of
  connected and automated vehicles at intersections and merging at highway
  on-ramps,'' \emph{IEEE Transactions on Intelligent Transportation Systems},
  vol.~18, no.~5, pp. 1066--1077, 2016.

\bibitem{jacobson1989real}
L.~N. Jacobson, K.~C. Henry, and O.~Mehyar, \emph{Real-time metering algorithm
  for centralized control}, 1989, no. 1232.

\bibitem{hourdakis2002evaluation}
J.~Hourdakis and P.~G. Michalopoulos, ``Evaluation of ramp control
  effectiveness in two twin cities freeways,'' \emph{Transportation Research
  Record}, vol. 1811, no.~1, pp. 21--29, 2002.

\bibitem{lin2019anti}
Y.~Lin, J.~McPhee, and N.~L. Azad, ``Anti-jerk on-ramp merging using deep
  reinforcement learning,'' in \emph{2020 IEEE Intelligent Vehicles Symposium
  (IV)}.\hskip 1em plus 0.5em minus 0.4em\relax IEEE, 2019, pp. 7--14.

\bibitem{cao2015cooperative}
W.~Cao, M.~Mukai, T.~Kawabe, H.~Nishira, and N.~Fujiki, ``Cooperative vehicle
  path generation during merging using model predictive control with real-time
  optimization,'' \emph{Control Engineering Practice}, vol.~34, pp. 98--105,
  2015.

\bibitem{MPCbook}
J.~B. Rawlings, D.~Q. Mayne, and M.~Diehl, \emph{Model predictive control:
  theory, computation, and design}.\hskip 1em plus 0.5em minus 0.4em\relax Nob
  Hill Publishing Madison, WI, 2017, vol.~2.

\bibitem{papageorgiou2003review}
M.~Papageorgiou, C.~Diakaki, V.~Dinopoulou, A.~Kotsialos, and Y.~Wang, ``Review
  of road traffic control strategies,'' \emph{Proceedings of the IEEE},
  vol.~91, no.~12, pp. 2043--2067, 2003.

\bibitem{papageorgiou2002freeway}
M.~Papageorgiou and A.~Kotsialos, ``Freeway ramp metering: An overview,''
  \emph{IEEE transactions on intelligent transportation systems}, vol.~3,
  no.~4, pp. 271--281, 2002.

\bibitem{papamichail2008traffic}
I.~Papamichail and M.~Papageorgiou, ``Traffic-responsive linked ramp-metering
  control,'' \emph{IEEE Transactions on Intelligent Transportation Systems},
  vol.~9, no.~1, pp. 111--121, 2008.

\bibitem{lubars2020combining}
J.~Lubars, H.~Gupta, A.~Raja, R.~Srikant, L.~Li, and X.~Wu, ``Combining
  reinforcement learning with model predictive control for on-ramp merging,''
  \emph{arXiv preprint arXiv:2011.08484}, 2020.

\bibitem{lillicrap2015continuous}
T.~P. Lillicrap, J.~J. Hunt, A.~Pritzel, N.~Heess, T.~Erez, Y.~Tassa,
  D.~Silver, and D.~Wierstra, ``Continuous control with deep reinforcement
  learning,'' \emph{arXiv preprint arXiv:1509.02971}, 2015.

\bibitem{wangmulti}
J.~Wang, T.~Shi, Y.~Wu, L.~Miranda-Moreno, and L.~Sun, ``Multi-agent graph
  reinforcement learning for connected automated driving.''

\bibitem{palanisamy2020multi}
P.~Palanisamy, ``Multi-agent connected autonomous driving using deep
  reinforcement learning,'' in \emph{2020 International Joint Conference on
  Neural Networks (IJCNN)}.\hskip 1em plus 0.5em minus 0.4em\relax IEEE, 2020,
  pp. 1--7.

\bibitem{kaushik2018parameter}
M.~Kaushik, K.~M. Krishna \emph{et~al.}, ``Parameter sharing reinforcement
  learning architecture for multi agent driving behaviors,'' \emph{arXiv
  preprint arXiv:1811.07214}, 2018.

\bibitem{ha2020leveraging}
P.~Y.~J. Ha, S.~Chen, J.~Dong, R.~Du, Y.~Li, and S.~Labi, ``Leveraging the
  capabilities of connected and autonomous vehicles and multi-agent
  reinforcement learning to mitigate highway bottleneck congestion,''
  \emph{arXiv preprint arXiv:2010.05436}, 2020.

\bibitem{bhalla2020deep}
S.~Bhalla, S.~G. Subramanian, and M.~Crowley, ``Deep multi agent reinforcement
  learning for autonomous driving,'' in \emph{Canadian Conference on Artificial
  Intelligence}.\hskip 1em plus 0.5em minus 0.4em\relax Springer, 2020, pp.
  67--78.

\bibitem{dong2020drl}
J.~Dong, S.~Chen, P.~Y.~J. Ha, Y.~Li, and S.~Labi, ``A drl-based multiagent
  cooperative control framework for cav networks: a graphic convolution q
  network,'' \emph{arXiv preprint arXiv:2010.05437}, 2020.

\bibitem{yu2019distributed}
C.~Yu, X.~Wang, X.~Xu, M.~Zhang, H.~Ge, J.~Ren, L.~Sun, B.~Chen, and G.~Tan,
  ``Distributed multiagent coordinated learning for autonomous driving in
  highways based on dynamic coordination graphs,'' \emph{IEEE Transactions on
  Intelligent Transportation Systems}, vol.~21, no.~2, pp. 735--748, 2019.

\bibitem{mnih2016asynchronous}
V.~Mnih, A.~P. Badia, M.~Mirza, A.~Graves, T.~Lillicrap, T.~Harley, D.~Silver,
  and K.~Kavukcuoglu, ``Asynchronous methods for deep reinforcement learning,''
  in \emph{International conference on machine learning}, 2016, pp. 1928--1937.

\bibitem{mnih2013playing}
V.~Mnih, K.~Kavukcuoglu, D.~Silver, A.~Graves, I.~Antonoglou, D.~Wierstra, and
  M.~Riedmiller, ``Playing atari with deep reinforcement learning,''
  \emph{arXiv preprint arXiv:1312.5602}, 2013.

\bibitem{chu2019multi}
T.~Chu, J.~Wang, L.~Codec{\`a}, and Z.~Li, ``Multi-agent deep reinforcement
  learning for large-scale traffic signal control,'' \emph{IEEE Transactions on
  Intelligent Transportation Systems}, vol.~21, no.~3, pp. 1086--1095, 2019.

\bibitem{berner2019dota}
OpenAI, :, C.~Berner, G.~Brockman, B.~Chan, V.~Cheung, P.~Dębiak, C.~Dennison,
  D.~Farhi, Q.~Fischer, S.~Hashme, C.~Hesse, R.~Józefowicz, S.~Gray,
  C.~Olsson, J.~Pachocki, M.~Petrov, H.~P. d.~O.~Pinto, J.~Raiman, T.~Salimans,
  J.~Schlatter, J.~Schneider, S.~Sidor, I.~Sutskever, J.~Tang, F.~Wolski, and
  S.~Zhang, ``Dota 2 with large scale deep reinforcement learning,'' 2019.

\bibitem{naderializadeh2021resource}
N.~Naderializadeh, J.~Sydir, M.~Simsek, and H.~Nikopour, ``Resource management
  in wireless networks via multi-agent deep reinforcement learning,''
  \emph{IEEE Transactions on Wireless Communications}, 2021.

\bibitem{chen2020powernet}
D.~Chen, Z.~Li, T.~Chu, R.~Yao, F.~Qiu, and K.~Lin, ``Powernet: Multi-agent
  deep reinforcement learning for scalable powergrid control,'' \emph{arXiv
  preprint arXiv:2011.12354}, 2020.

\bibitem{tan1993multi}
M.~Tan, ``Multi-agent reinforcement learning: Independent vs. cooperative
  agents,'' in \emph{Proceedings of the tenth international conference on
  machine learning}, 1993, pp. 330--337.

\bibitem{lowe2017multi}
R.~Lowe, Y.~I. Wu, A.~Tamar, J.~Harb, O.~P. Abbeel, and I.~Mordatch,
  ``Multi-agent actor-critic for mixed cooperative-competitive environments,''
  in \emph{Advances in neural information processing systems}, 2017, pp.
  6379--6390.

\bibitem{lin2018efficient}
K.~Lin, R.~Zhao, Z.~Xu, and J.~Zhou, ``Efficient large-scale fleet management
  via multi-agent deep reinforcement learning,'' in \emph{Proceedings of the
  24th ACM SIGKDD International Conference on Knowledge Discovery \& Data
  Mining}, 2018, pp. 1774--1783.

\bibitem{terry2021revisiting}
J.~K. Terry, N.~Grammel, A.~Hari, L.~Santos, and B.~Black, ``Revisiting
  parameter sharing in multi-agent deep reinforcement learning,'' 2021.

\bibitem{schulman2017proximal}
J.~Schulman, F.~Wolski, P.~Dhariwal, A.~Radford, and O.~Klimov, ``Proximal
  policy optimization algorithms,'' \emph{arXiv preprint arXiv:1707.06347},
  2017.

\bibitem{wu2017scalable}
Y.~Wu, E.~Mansimov, S.~Liao, R.~Grosse, and J.~Ba, ``Scalable trust-region
  method for deep reinforcement learning using kronecker-factored
  approximation,'' \emph{arXiv preprint arXiv:1708.05144}, 2017.

\bibitem{elsayed2021safe}
I.~Elsayed-Aly, S.~Bharadwaj, C.~Amato, R.~Ehlers, U.~Topcu, and L.~Feng,
  ``Safe multi-agent reinforcement learning via shielding,'' \emph{arXiv
  preprint arXiv:2101.11196}, 2021.

\bibitem{cai2021safe}
Z.~Cai, H.~Cao, W.~Lu, L.~Zhang, and H.~Xiong, ``Safe multi-agent reinforcement
  learning through decentralized multiple control barrier functions,''
  \emph{arXiv preprint arXiv:2103.12553}, 2021.

\bibitem{hausknecht2015deep}
M.~Hausknecht and P.~Stone, ``Deep recurrent q-learning for partially
  observable mdps,'' \emph{arXiv preprint arXiv:1507.06527}, 2015.

\bibitem{li2017explicit}
N.~Li, H.~Chen, I.~Kolmanovsky, and A.~Girard, ``An explicit decision tree
  approach for automated driving,'' in \emph{Dynamic Systems and Control
  Conference}, vol. 58271.\hskip 1em plus 0.5em minus 0.4em\relax American
  Society of Mechanical Engineers, 2017, p. V001T45A003.

\bibitem{chen2020autonomous}
D.~Chen, L.~Jiang, Y.~Wang, and Z.~Li, ``Autonomous driving using safe
  reinforcement learning by incorporating a regret-based human lane-changing
  decision model,'' in \emph{2020 American Control Conference (ACC)}.\hskip 1em
  plus 0.5em minus 0.4em\relax IEEE, 2020, pp. 4355--4361.

\bibitem{transportation2011policy}
T.~Officials, \emph{A Policy on Geometric Design of Highways and Streets,
  2011}.\hskip 1em plus 0.5em minus 0.4em\relax AASHTO, 2011.

\bibitem{thiemann2008estimating}
C.~Thiemann, M.~Treiber, and A.~Kesting, ``Estimating acceleration and
  lane-changing dynamics from next generation simulation trajectory data,''
  \emph{Transportation Research Record}, vol. 2088, no.~1, pp. 90--101, 2008.

\bibitem{ayres2001preferred}
T.~Ayres, L.~Li, D.~Schleuning, and D.~Young, ``Preferred time-headway of
  highway drivers,'' in \emph{ITSC 2001. 2001 IEEE Intelligent Transportation
  Systems. Proceedings (Cat. No. 01TH8585)}.\hskip 1em plus 0.5em minus
  0.4em\relax IEEE, 2001, pp. 826--829.

\bibitem{treiber2000congested}
M.~Treiber, A.~Hennecke, and D.~Helbing, ``Congested traffic states in
  empirical observations and microscopic simulations,'' \emph{Physical review
  E}, vol.~62, no.~2, p. 1805, 2000.

\bibitem{kesting2007general}
A.~Kesting, M.~Treiber, and D.~Helbing, ``General lane-changing model mobil for
  car-following models,'' \emph{Transportation Research Record}, vol. 1999,
  no.~1, pp. 86--94, 2007.

\bibitem{polack2017kinematic}
P.~Polack, F.~Altch{\'e}, B.~d'Andr{\'e}a Novel, and A.~de~La~Fortelle, ``The
  kinematic bicycle model: A consistent model for planning feasible
  trajectories for autonomous vehicles?'' in \emph{2017 IEEE intelligent
  vehicles symposium (IV)}.\hskip 1em plus 0.5em minus 0.4em\relax IEEE, 2017,
  pp. 812--818.

\bibitem{wolpert2002optimal}
D.~H. Wolpert and K.~Tumer, ``Optimal payoff functions for members of
  collectives,'' in \emph{Modeling complexity in economic and social
  systems}.\hskip 1em plus 0.5em minus 0.4em\relax World Scientific, 2002, pp.
  355--369.

\bibitem{wang2020shapley}
J.~Wang, Y.~Zhang, T.-K. Kim, and Y.~Gu, ``Shapley q-value: A local reward
  approach to solve global reward games,'' in \emph{Proceedings of the AAAI
  Conference on Artificial Intelligence}, vol.~34, no.~05, 2020, pp.
  7285--7292.

\bibitem{sutton2018reinforcement}
R.~S. Sutton and A.~G. Barto, \emph{Reinforcement learning: An
  introduction}.\hskip 1em plus 0.5em minus 0.4em\relax MIT press, 2018.

\bibitem{bagnell2005local}
D.~Bagnell and A.~Ng, ``On local rewards and scaling distributed reinforcement
  learning,'' \emph{Advances in Neural Information Processing Systems},
  vol.~18, pp. 91--98, 2005.

\bibitem{williams1992simple}
R.~J. Williams, ``Simple statistical gradient-following algorithms for
  connectionist reinforcement learning,'' \emph{Machine learning}, vol.~8, no.
  3-4, pp. 229--256, 1992.

\bibitem{huang2020closer}
S.~Huang and S.~Onta{\~n}{\'o}n, ``A closer look at invalid action masking in
  policy gradient algorithms,'' \emph{arXiv preprint arXiv:2006.14171}, 2020.

\bibitem{chou2009feasibility}
C.-M. Chou, C.-Y. Li, W.-M. Chien, and K.-c. Lan, ``A feasibility study on
  vehicle-to-infrastructure communication: Wifi vs. wimax,'' in \emph{2009
  tenth international conference on mobile data management: systems, services
  and middleware}.\hskip 1em plus 0.5em minus 0.4em\relax IEEE, 2009, pp.
  397--398.

\bibitem{highway-env}
E.~Leurent, ``An environment for autonomous driving decision-making,''
  \url{https://github.com/eleurent/highway-env}, 2018.

\bibitem{kaushik2018overtaking}
M.~Kaushik, V.~Prasad, K.~M. Krishna, and B.~Ravindran, ``Overtaking maneuvers
  in simulated highway driving using deep reinforcement learning,'' in
  \emph{2018 ieee intelligent vehicles symposium (iv)}.\hskip 1em plus 0.5em
  minus 0.4em\relax IEEE, 2018, pp. 1885--1890.

\bibitem{cao2013two}
W.~Cao, M.~Mukai, and T.~Kawabe, ``Two-dimensional merging path generation
  using model predictive control,'' \emph{Artificial Life and Robotics},
  vol.~17, no. 3-4, pp. 350--356, 2013.

\bibitem{saxena2020driving}
D.~M. Saxena, S.~Bae, A.~Nakhaei, K.~Fujimura, and M.~Likhachev, ``Driving in
  dense traffic with model-free reinforcement learning,'' in \emph{2020 IEEE
  International Conference on Robotics and Automation (ICRA)}.\hskip 1em plus
  0.5em minus 0.4em\relax IEEE, 2020, pp. 5385--5392.

\bibitem{alshiekh2018safe}
M.~Alshiekh, R.~Bloem, R.~Ehlers, B.~K{\"o}nighofer, S.~Niekum, and U.~Topcu,
  ``Safe reinforcement learning via shielding,'' in \emph{Proceedings of the
  AAAI Conference on Artificial Intelligence}, vol.~32, no.~1, 2018.

\bibitem{garcia2015comprehensive}
J.~Garc{\i}a and F.~Fern{\'a}ndez, ``A comprehensive survey on safe
  reinforcement learning,'' \emph{Journal of Machine Learning Research},
  vol.~16, no.~1, pp. 1437--1480, 2015.

\bibitem{zhao2020sim}
W.~Zhao, J.~P. Queralta, and T.~Westerlund, ``Sim-to-real transfer in deep
  reinforcement learning for robotics: a survey,'' in \emph{2020 IEEE Symposium
  Series on Computational Intelligence (SSCI)}.\hskip 1em plus 0.5em minus
  0.4em\relax IEEE, 2020, pp. 737--744.

\end{thebibliography}
\bibliographystyle{IEEEtran}

\end{document}